\documentclass[prb, twocolumn, showpacs, amsmath, amssymb,superscriptaddress]{revtex4-1}
\usepackage{amssymb,amsfonts,amsmath} 
\usepackage{graphicx}
\usepackage{color}
\usepackage[colorlinks=true, citecolor=blue, linkcolor=blue, urlcolor=blue]{hyperref}
\usepackage{longtable}
\usepackage{bbm}
\usepackage{enumitem}
\usepackage{bm}
\usepackage{braket}
\usepackage{comment}
\usepackage{cancel}
\usepackage{nicefrac}

\begin{document}

\title{Anti-Poiseuille Flow: Increased Vortex Velocity at Superconductor Edges}

\author{T. Okugawa}
\affiliation{Institut f\"ur Theorie der Statistischen Physik, RWTH Aachen, 
52056 Aachen, Germany and JARA - Fundamentals of Future Information Technology.}
\author{A. Benyamini}
\affiliation{Department of Physics, Columbia University, New York, NY 10027, USA}
\author{A. J. Millis}
\affiliation{Department of Physics, Columbia University, New York, NY 10027, USA}
\affiliation{Center for Computational Quantum Physics, Flatiron Institute, 162 5th Avenue, New York, NY 10010 USA}
\author{D. M. Kennes}
\email{Dante.Kennes@rwth-aachen.de}
\affiliation{Institut f\"ur Theorie der Statistischen Physik, RWTH Aachen, 
52056 Aachen, Germany and JARA - Fundamentals of Future Information Technology.}
\affiliation{Max Planck Institute for the Structure and Dynamics of Matter, Center for Free Electron Laser Science, Luruper Chaussee 149, 22761 Hamburg, Germany.}

\date{\today}

\begin{abstract}
Using the time-dependent Ginzburg Landau equations we study vortex motion driven by an applied current in two dimensional superconductors in the presence of a physical boundary. At smaller sourced currents the vortex lattice moves as a whole, with each vortex moving at the same velocity. At larger sourced current, vortex motion is organized into channels, with vortices in channels nearer to the sample edges moving faster than those farther away from sample edges, opposite to the Poiseuille flow of basic hydrodynamics where the velocity is lowest at the boundaries. At intermediate currents, a stick-slip motion of the vortex lattice occurs in which vortices in the channel at the boundary break free from the Abrikosov lattice, accelerate, move past their neighbors and then slow down 
and reattach to the vortex lattice at which point the stick-slip process starts over. These effects could be observed experimentally, e.g. using fast scanning microscopy techniques.
\end{abstract}
\maketitle

\section{Introduction}
In the presence of an applied magnetic field a type-II superconductor may exhibit a vortex state, characterized by a certain density of order parameter singularities (vortices). In equilibrium, and in the absence of important disorder effects, the vortices generically form an ordered ``Abrikosov lattice".\cite{Abrikosov1957} An applied current produces a force on the vortices, and the resulting vortex motion leads to dissipation,\cite{Abrikosov1957, Bardeen1965, Tinkham} which is one of the main concerns for many applications such as superconducting qubits and superconducting digital memory,\cite{Devoret_2013, met9101022} high-field magnets,\cite{Si_2013} THz radiation sources,\cite{Welp_2013} or resonant cavities in particle accelerator.\cite{Padamsee_2001, Gurevich2008} A deepened understanding of the nature of vortex motion may enhance device performances enabling many intriguing applications.\cite{Veshchunov_2016, Madan_2018, Rouco2019, Tamireaau2019}

Vortex structures driven by an external current in systems without pinning sites have been studied.\cite{Embon2017, Dobrovolskiy_2020, Vodolazov2007, Vodolazov2013, Grimaldi2015} Of particular interest is the ''phase slip line state"\cite{Berdiyorov2014, Andronov1993, Sivakov2003} where a set of parallel vortex rows are created and each vortex follows in its own "assigned" channel. This phenomena originates from an effective attraction to the other vortices due to a suppression of the order parameter created behind each moving vortex.\cite{Keizer2006, Berdiyorov2009, Silhanek2010} This mechanism was analysed in the context of the viscosity and the instability of the vortex motion using both Bardeen-Stephen and Larkin-Ovchinnikov theory.\cite{Bardeen1965, Larkin1975, Vodolazov2007} However, finite size effects, 
in particular boundaries which are always present in real superconducting devices, have received little attention in this context so far. 

In this work, we use time-dependent Ginzburg-Landau (TGDL) theory\cite{Gorkov1959, Schmid1966, Cyrot1973} to investigate the flux flow properties of clean two dimensional superconductors driven by externally sourced currents,  in finite geometries. We find a low current regime where the vortex lattice moves as a whole, a high current ``anti-Poiseuille" regime in which vortices near the sample boundaries move faster, and an intermediate regime characterized by a ``stick-slip" behavior. We show that our results can be understood in terms of a Bardeen-Stephen\cite{Bardeen1965} analysis where the reduced order parameter close to the edge provides a lower viscosity for the vortex motion. 

The rest of this paper is organized as follows. In section \ref{formalism}, we present our model and the computational method adopted for this paper. In section \ref{result} and \ref{interpretation}, we show the main results of our calculations and the corresponding discussion, respectively. Finally, section \ref{conclusion} concludes the paper with a summary of results and discussion of future work.

\section{Formalism}
\label{formalism}
We use TDGL equations to describe the dynamics of a complex superconducting order parameter $\Delta=|\Delta|e^{\mathrm{i}\phi}$ in a two dimensional strip geometry in the presence of both an external current created by a source and drain of particles whose strength is denoted by $Q$ and an external magnetic field $\bm{B}$ directed perpendicular to the superconducting film. The electromagnetic field is expressed by the vector potential $\bm{A}$ and the scalar potential $\Theta$. We denote the charge density as $\rho$ and current density as $\bm{J}$ and choose unit as $\hbar=c=e=1$. The equations are:\cite{Tinkham, Kennes, Benyamini_2019, Benyamini_2020}
\begin{align}
&\frac{1}{D} \left(\partial_{t} +2\mathrm{i} \Psi \right)\Delta = \frac{1}{\xi^2 \beta} \Delta [\alpha - \beta |\Delta |^2] \notag \\
&+ [ \bm{ \nabla} - 2 \mathrm{i}\bm{A}  ]^2 \Delta,  \label{TDGL} \\
&\bm{J}=\sigma [ -\bm{\nabla}\Psi - \partial_{t}\bm{A}  ] + \sigma \tau_{s} \Re[\Delta^{\ast} \left( -  \mathrm{i} \bm{\nabla} - 2 \bm{A}      \right)\Delta ], \label{current} \\
&\rho = \frac{\Psi - \Theta}{4 \pi \lambda_{TF}^2}, \label{final} \\
&\partial_{t} \rho + \bm{\nabla} \cdot \bm{J}=Q, \label{continuity}\\
&\bm{\nabla}^2 \Theta = -4 \pi \rho. \label{poisson}
\end{align}
Here $D$ is the normal state diffusion constant, $\Psi$ is the electrochemical potential per electron charge, $\xi=\sqrt{6D/\tau_{s}}$, and the superconducting coherent length is given as $\xi_{0} = \xi/\sqrt{\alpha/\beta}$, where $\tau_{s}$ is the spin-flip scattering time. $\alpha$ and $\beta$ are system dependent constants, which set the magnitude of the order parameter. 
$\sigma$ and $\lambda_{TF}$ are the normal state conductivity and the Thomas-Fermi static charge screening length, respectively. 
We measure length in units of $\xi$ and time in units $\xi^2/D$ (since $\xi$ is the unit of length, we write this as $D^{-1}$). We choose the parameters as follows: $\alpha=\beta=1$, $\tau_s=6 D/\xi^2$, $\lambda_{TF}/\xi=1$, $\sigma/(D/\xi^2)=1$, and set the length and width of the two dimensional strip to be $L=50 \xi$. We confirmed that this particular parameter choice does not affect the general conclusions of this paper. 
We choose the Coulomb gauge $\nabla\cdot A=0$~\footnote{However, the Gauge function $\chi$ cannot be fixed uniquely only by this condition for the finite element method, i.e $\chi$ can only be determined up to a constant. Thus, we introduce another condition; $\int_S \Psi dS=0$, where $S$ is the entire sample domain.\cite{Qiang1994, fan2019}}.

We study a system that is periodic in $y$ and with open superconductor-vacuum boundary conditions in $x$:\cite{Vodolazov2007, Carapella_2011, Sabatino_2011}
\begin{align}
\Delta(y+L)&=\Delta(y),\label{ybc} \\
\left(-\mathrm{i} \nabla_x - A_x \right) \Delta|_{x=0, ~L} &= 0,  \label{no_bound_current}\\
\left(-(\bm{\nabla}\Psi)_x - \partial_{t}A_x  \right)|_{x=0,~L}  &= 0.  \label{no_normal_current}
\end{align}
The advantage of the periodic boundary condition is that it avoids the need to consider complications, irrelevant here, related to creation and destruction of vortices at sample edges.~\footnote{When vortices move out of the finite sample, critical force which overcomes any barrier is needed. For a large enough slab, such an effect will be irrelevant since the collective force of the current over all vortices is much larger than any barrier.}

\begin{figure}[t]
\begin{center} 
\includegraphics[width=0.49\textwidth, angle=-0]{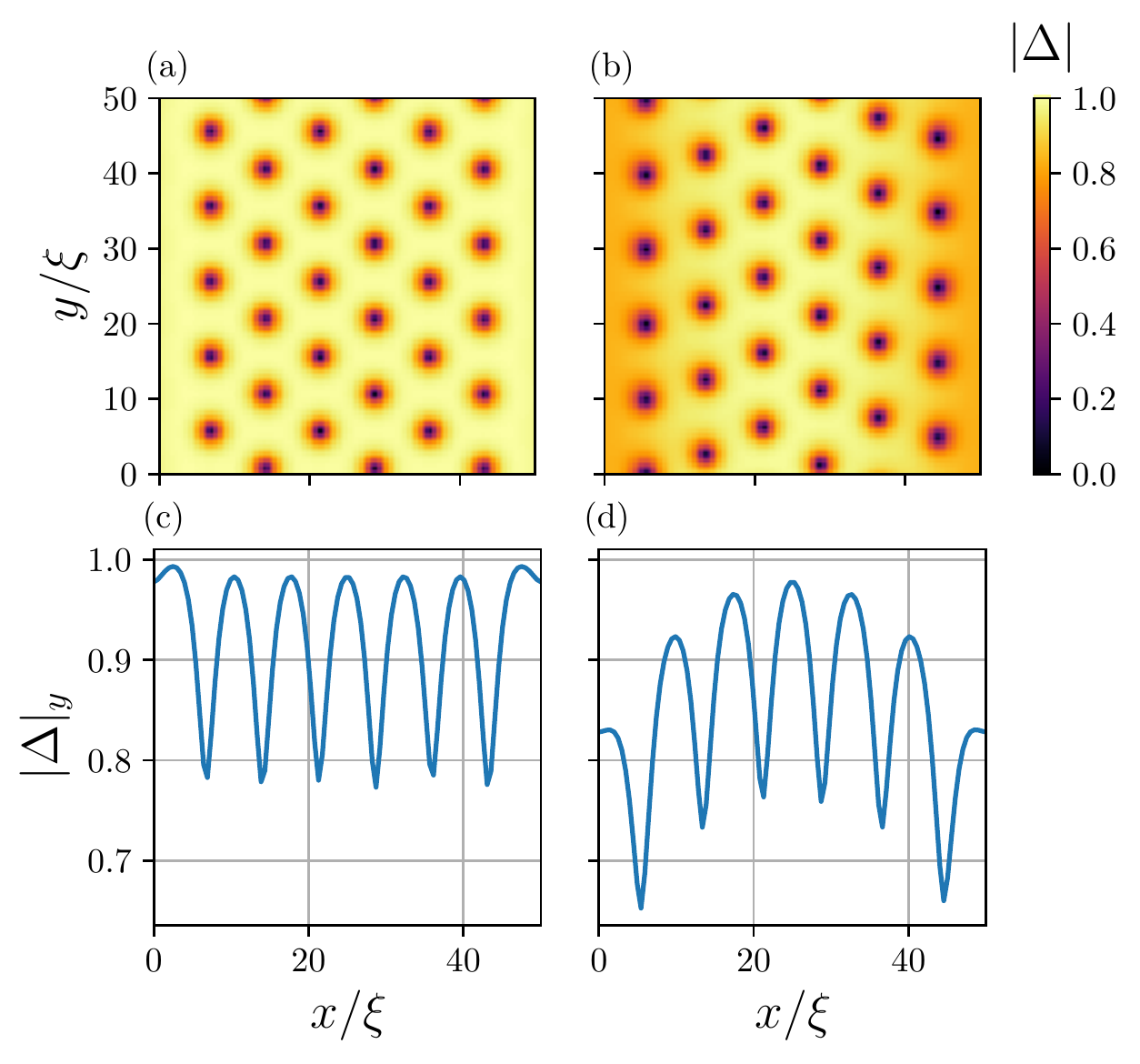}
\caption{False color representation of the magnitude of the superconducting order parameter $|\Delta|$ in the upper panel and the corresponding $y$-averaged order parameter, denoted as $|\Delta|_{y}$ against $x$ in the lower panel, calculated with $\xi^2 B=0.043$ for zero applied current ($Q/(D^2/\xi^6)= 0$) in (a), (c) and nonzero applied current ($Q/(D^2/\xi^6)= 0.55$) in (b), (d). The configuration depicted in (b) is a snapshot showing one instant in time of a dynamic state. 
The corresponding video is available in the SM as video14.\cite{SM}
} 
\label{fig:edge_0}
\end{center}
\end{figure}

We consider very thin films where the sample thickness is much less than the London penetration depth so that the magnetic field can be taken to be spatially uniform 
and described by the Coulomb-gauge vector potential $\bm{A}=(0, B(x-L/2), 0)$.
We choose magnetic field values such that the number of vortices is commensurate with the system size so that each vertical line of vortices in Fig.~\ref{fig:edge_0} has the same number of vortices to avoid additional complications that arise when there is an additional vortex which could go into any one of the channels. 

The external current is introduced by the source term $Q$ in Eq.~\eqref{continuity}, which defines source and drain regions of width $\xi$ over the entire length of the open boundaries. Specifically we take $Q$ to be independent of $y$ and $Q(x, y) \neq 0~\text{only for}~0<x<\xi~\text{and}~L-\xi<x<L $. Thus, this current flow is homogeneous in the entire sample (except the source and drain sections). 

\begin{figure*}[t]
\begin{center} 
\includegraphics[width=0.95\textwidth, angle=-0]{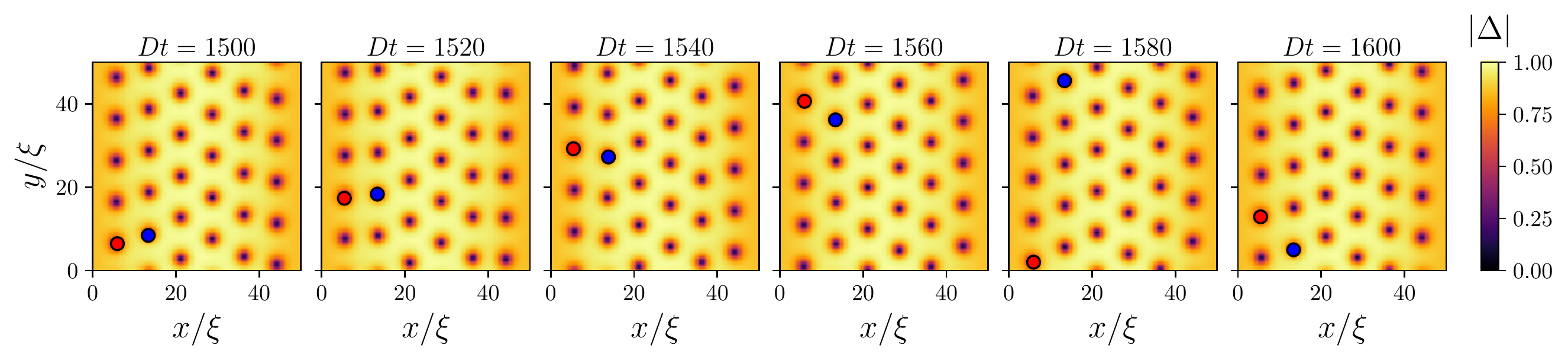}
\caption{Series of false color plots of the magnitude of the order parameter $|\Delta|$ for $\xi^2 B=0.043$ and $Q/(D^2/\xi^6)=0.55$ at different times. The same two vortices are labeled by red and blue dots for each time to highlight that the vortices close to the edge (red dot) move faster than the ones in the bulk (blue dot). Each plot is a snapshot showing instances in time which are separated by $20D^{-1}$ from left to right and the initial time is $Dt=1500$. The corresponding video is available in SM as video14.\cite{SM}}
\label{fig:snap}
\end{center}
\end{figure*}

The simulations are performed using a finite element method in space implemented in FEniCS\cite{LoggMardalEtAl2012a} and we discretize the time derivative by a finite difference approximation 
in which the order parameter is updated in time steps of $D\Delta t$ = 0.5. We have verified that decreasing the time step does not change the results. 
The initial conditions (at time $t=0$) of all variables except $\Delta$ are set to be zero, while $\Delta(x, y, t=0)$ is chosen to describe the superconducting state with some random fluctuation over the entire 2D sample.~\footnote{Specifically, both real and imaginary part of the complex order parameter $\Re[\Delta]$ and $\Im[\Delta]$ are set to be $\sqrt{0.5}+\delta$, where $\delta$ is chosen randomly from a uniform distribution $\delta \in (-0.01, 0.01]$. However, since we concentrate on analyzing the steady state none of the general conclusions are affected by the initial conditions.}  

In our calculation protocol we first set the current to zero and evolve the system; once a stable vortex lattice is formed we introduce the current, and then analyse the flow after a steady state is reached. To be specific, non-zero $Q$ is introduced at $Dt=1000$, when a rigid vortex structure has already been formed and an analysis of vortex flow created by the sourced current is performed after $Dt=1500$.

\section{Results}
\label{result}

In this section we present the main features of our results. Before doing so we summarize the expected physics. In a magnetic field, the superconducting order parameter of a thin film exhibits vortices, points at which the order parameter vanishes and around which there is a quantized circulation of the superconducting phase. The density of vortices is set by the applied field and the energy associated with superposing the quantized circulations leads to vortex-vortex interactions that are repulsive, and logarithmic at long scales; the resulting  equilibrium configuration is a triangular ``Abrikosov lattice" with the lattice vectors parallel or at angles of $\pm \pi/3$ to the open boundaries at $x=0,L$ (shown in Fig.~\ref{fig:edge_0} (a) as an example). In this state, if the order parameter amplitude is averaged over $y$ and then plotted against $x$, vortex rows corresponding to equally spaced minima in the order parameter amplitude are evident (shown in Fig.~\ref{fig:edge_0} (c) as an example). In the presence of an applied current $\bm{J}$ a vortex is accelerated by a Lorentz force $\bm{F}=\bm{J}\times \bm{B}/c$ directed perpendicular to the current and the magnetic field, and experiences a viscosity $\eta$, which is within Bardeen-Stephen picture given by\cite{Tinkham}
\begin{equation}
\eta=\Phi_0\sigma \frac{H_{c_2}}{c^2}\sim \Delta^{2},
\label{eta}
\end{equation}
where $\sigma$ is the conductivity appearing in the TDGL equations and $\Phi_0$ is the flux quantum. $H_{c2}$ is the upper critical field which is proportional to the inverse of the square of the superconducting order parameter in the clean limit studied here. The system we study has a translation invariance in the $y$ direction so that in the presence of a current one possible solution is a uniformly translating vortex lattice in which the velocity is determined from the Lorentz force and the viscosity. 
As we shall see, in our system, other solutions are possible, corresponding to non-uniform vortex motion. In all of our investigations, the non-uniformly moving state exhibits the same row structure as in the equilibrium and uniformly moving states, but vortices in different rows may move with different velocities.
\subsection{Equilibrium}
\label{sub:Equilibrium}

First, we perform simulations without an externally sourced current, $Q=0$. We initialize our simulation as described in section \ref{formalism} 
and allow the system to evolve to a time independent state. The time evolution takes around 600$D^{-1}$. 
The result is shown in Fig.~\ref{fig:edge_0} (a), which presents a false-color plot of the magnitude of the order parameter $|\Delta|$ against position. As expected, an Abrikosov vortex lattice is formed, with the lattice aligned to the boundaries. The vortex positions are visible as regions of very small order parameter amplitude, and the row structure alluded to above is evident. The periodic boundary conditions along $y$ mean that there is a continuous family of equilibrium solutions translated by arbitrary amounts along $y$.    

Remarkably, in the equilibrium case there are essentially no boundary effects except for the alignment of the lattice and that the vortices avoid the sample boundaries because the supercurrent is reflected at the boundary. The order parameter amplitude in the region away from the vortices shows no significant variation with $x$ as in Fig.~\ref{fig:edge_0} (c), where we plot the $y-$averaged (flow direction) order parameter of panel (a), denoted as $|{\Delta}|_{y}$. 

\subsection{Nonzero current drive}
At small applied currents our numerically computed solution is just a uniformly translating vortex lattice. However above a critical current the lattice structure is somewhat disrupted. The parallel chain structure remains, but the vortices in the chain closest to the sample edge flow at a higher velocity.  This is illustrated in Fig. ~\ref{fig:snap} which shows a series of false color plots of the order parameter amplitude presented at time intervals $20D^{-1}$ apart. Two vortices, one in the left most row and one in the second row from the left are highlighted by red and blue color respectively. The difference in velocities is evident.
\begin{figure}[t!]
\begin{center} 
\includegraphics[width=0.45\textwidth, angle=-0]{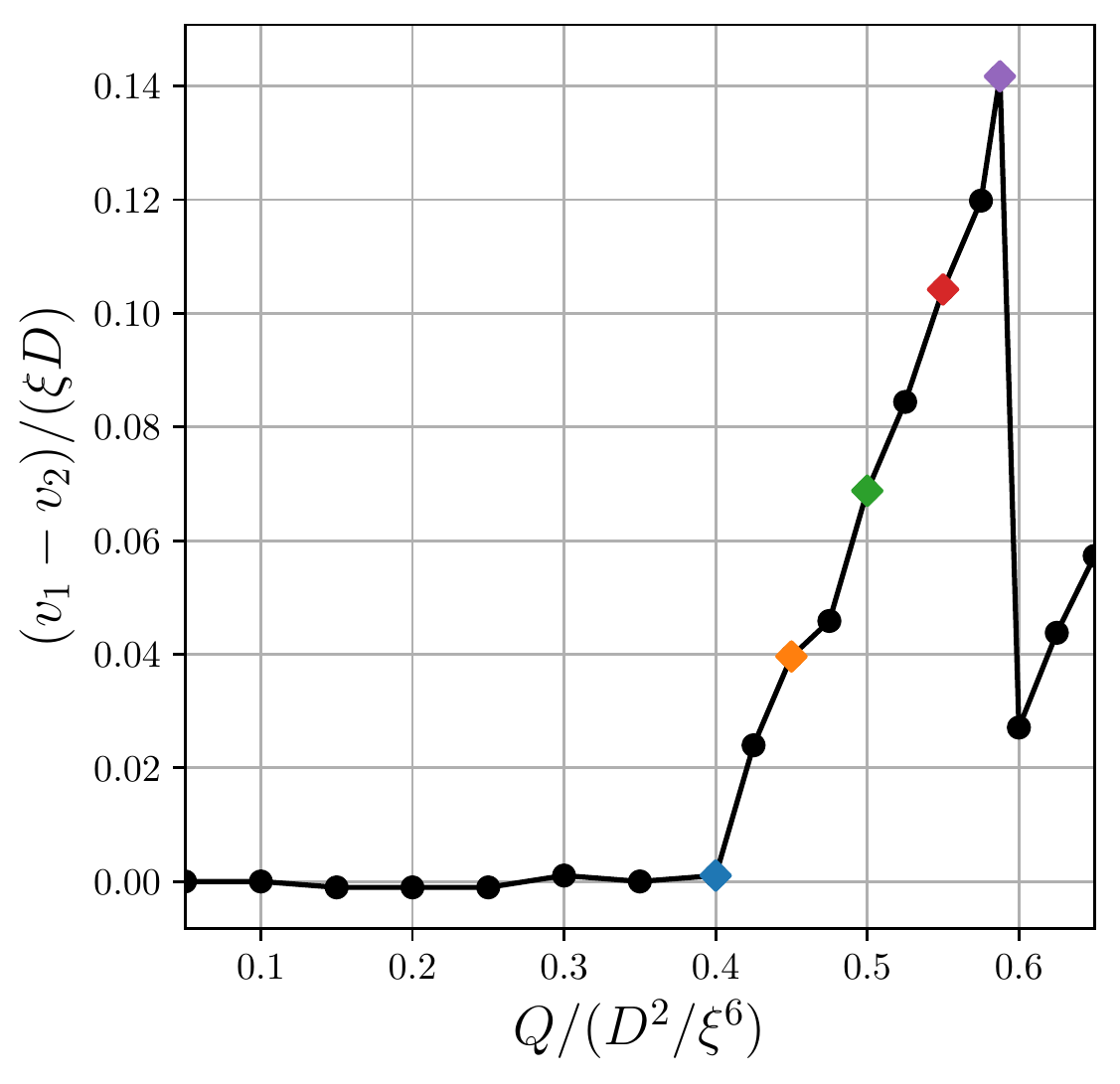}
\caption{Differences of the time averaged velocities of two tagged particles, $v_{1}$ of a vortex in the left most row and $v_2$ of a vortex in the adjacent row (e.g. the red and blue vortices in Fig.~\ref{fig:snap}, respectively) as a function of the strength of the source term $Q$. The vortex speeds are extracted from the simulations averaging from $Dt=1500$ to $Dt=1980$. The color of the diamonds in the figure corresponds to the parameters shown in Fig. \ref{fig:C1vsJ}. All corresponding videos are available in the SM.\cite{SM} (see also TABLE \ref{table:video} in Appendix \ref{appendix:video}).}
\label{fig:vel1}
\end{center}
\end{figure}

To quantify the dependence of the differential vortex velocity on the magnitude of the source term $Q$ in Eq. \eqref{continuity}, we present in Fig.~\ref{fig:vel1} a plot of the difference of the time averaged velocities of two tagged particles, $v_{1}$ of a vortex in the left most row and $v_2$ of a vortex in the adjacent row (e.g. the red and blue vortices in Fig.~\ref{fig:snap}, respectively) as a function of $Q$. We see that the difference in velocities remains zero up to a critical current $Q\approx 0.4D^2/\xi^6$. For larger $Q$ the time averaged velocity difference increases linearly with $Q-Q_c$ until at a yet larger $Q\approx 0.6D^2/\xi^6$ the system is reset: the number of vortex channels changes from $6$ to $4$ due to a strong suppression of the order parameter close to the open boundary (see the SM video17, 18, 19~\cite{SM}). In the newly established $4$ channel structure, the same characteristic as for the intermediate regime $0.4 < Q/(D^2/\xi^6) < 0.6$ is observed as shown in Fig.~\ref{fig:vel1} by a linear increase as a function of $Q/(D^2/\xi^6)$ after the jump in the curve.

\section{Interpretation}
\label{interpretation}
\begin{figure}[t!]
\begin{center} 
\includegraphics[width=0.45\textwidth, angle=-0]{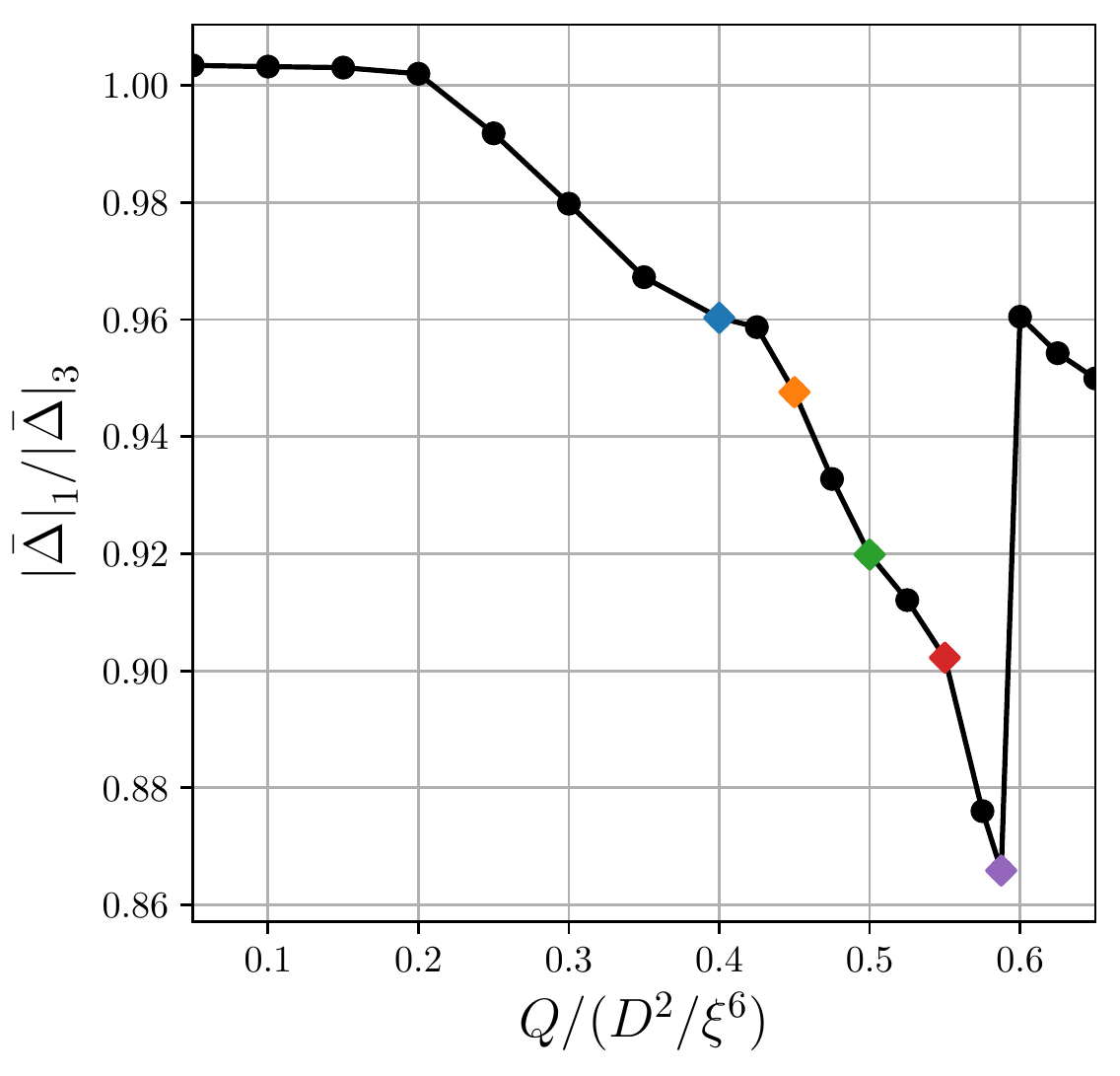}
\caption{Ratio of the averaged order parameter amplitude between $\bar{|{\Delta}|}_{1}$ in the left most row (edge) and $\bar{|{\Delta}|}_{3}$ in the middle left row (bulk) as a function of the strength of the source term $Q$. The $x-$dependent order parameter $\bar{|\Delta|}$ are calculated by averaging over the $y-$direction and over the same time duration as for the vortex speed shown in Fig. \ref{fig:vel1}. The color of the diamonds in the figure corresponds to the parameters shown in Fig. \ref{fig:C1vsJ}. All corresponding videos are available in the SM.\cite{SM} (see also TABLE \ref{table:video} in Appendix \ref{appendix:video}). 
}
\label{fig:gapvsJ}
\end{center}
\end{figure}
We now argue that the velocity differential arises from a position dependence of the viscosity. We see from Fig.~\ref{fig:edge_0} (b), (d) that at high drive current the typical value of $|\Delta|$, $|\Delta|_y$ far from a vortex is smaller near the sample edge than in the middle, implying via Eq.~\eqref{eta} that the vortices closest to the sample edge experience a lower viscosity. As clearly seen from the comparison between Fig.~\ref{fig:edge_0} (c) and (d), in the presence of an externally sourced current, $|\Delta|_y$ in the region away from the vortices is reduced as it goes to the sample edge. To study the dependence of the reduction of the order parameter on the strength of the source term $Q$ in Eq. \eqref{continuity}, we define the time and y-direction averaged order parameter amplitude, denoted as $\bar{|{\Delta}|}$ in each vortex row and plot in Fig \ref{fig:gapvsJ} the ratio of $\bar{|{\Delta}|}$ assigned in the two rows, $\bar{|{\Delta}|}_{1}$ in the left most row and $\bar{|{\Delta}|}_{3}$ in middle left row. The ratio $\bar{|{\Delta}|}_{1}/\bar{|{\Delta}|}_{3}$ monotonously decreases both before and after $Q\approx 0.6D^2/\xi^6$ where the number of vortex channels changes from $6$ to $4$, which implies that the viscosity difference between the left most and middle left row increases as the sourced current increases within the systems of the same number of vortex channels. 

Now let us consider the equation of motion in $y$-direction for a vortex in the left-most row, supposing that the other vortices in the system are in an ordered, uniformly translating, state:
\begin{equation}
    F_{VL}(y)+\frac{J\Phi_0}{c}=\eta_1\dot{y}.
    \label{EOM}
\end{equation}
Here $F_{VL}$ is the force arising from all of the other vortices. The current term may be related to the velocity of the uniformly translating part of the vortex lattice, denoted $V$ as $\frac{J\Phi_0}{c}=\bar{\eta}V$, where $\bar{\eta}$ is the average viscosity relevant to these vortices. In the frame co-moving with the uniformly translating state Eq.~\eqref{EOM} becomes
\begin{equation}
    F_{VL}(y) +(\bar{\eta}-\eta_1)V=\eta_1\dot{y}.
    \label{EOMf}
\end{equation}
In other words, in the co-moving frame the vortices at the edge of the sample experience two forces--one set by the mean velocity and the viscosity difference, and one from the lattice of uniformly moving vortices. This latter force is periodic under translation by the vortex lattice vector and has a maximum value. At small current, both the difference in viscosity and the mean velocity are small, so this force is insufficient to overcome the force from the vortex lattice. The solution in the co-moving frame is then 
$y=0$ with $y$ shifted from its equilibrium position by an amount proportional to the force. As the current is increased, the $(\bar{\eta}-\eta_1)V$ term increases, the shift in mean position increases, and when the position exceeds the point of maximum lattice force the solution in the co-moving frame becomes time dependent.


\subsection{Anti-Poiseuille flow}
As is discussed above, in the non-uniformly moving state ($0.4 < Q/(D^2/\xi^6)$ in Fig. \ref{fig:C1vsJ}), the vortices move faster at the edge of the sample because the vortices closest to the edge experience a lower viscosity due to the reduction of the order parameter, shown in Eq. \eqref{eta}. To quantify the relation between the vortex velocity and the reduction of the order parameter, here we perform a simple analysis by neglecting vortex-vortex interaction and the corresponding force arising from it, $F_{VL}\approx 0$, which enables us to consider the dynamics of each vortex row independently, 
and may be reasonable in the large current regime of our present study. In this case, Eq. \eqref{EOM} can be applied to all vortex rows, which is given as $J\frac{\Phi_0}{c}=\eta \dot{y}$. Since the current $J$ is homogeneous over almost the entire two-dimensional sample in our simulations (except the source and drain sections), together with Eq. \eqref{eta}, we can get a simple prediction for the vortex speed and the order parameter:

\begin{equation}
\dot{y} \propto \Delta^{-2}\label{eq:predv}.
\end{equation}

\begin{figure}[t!]
\begin{center} 
\includegraphics[width=0.49\textwidth, angle=-0]{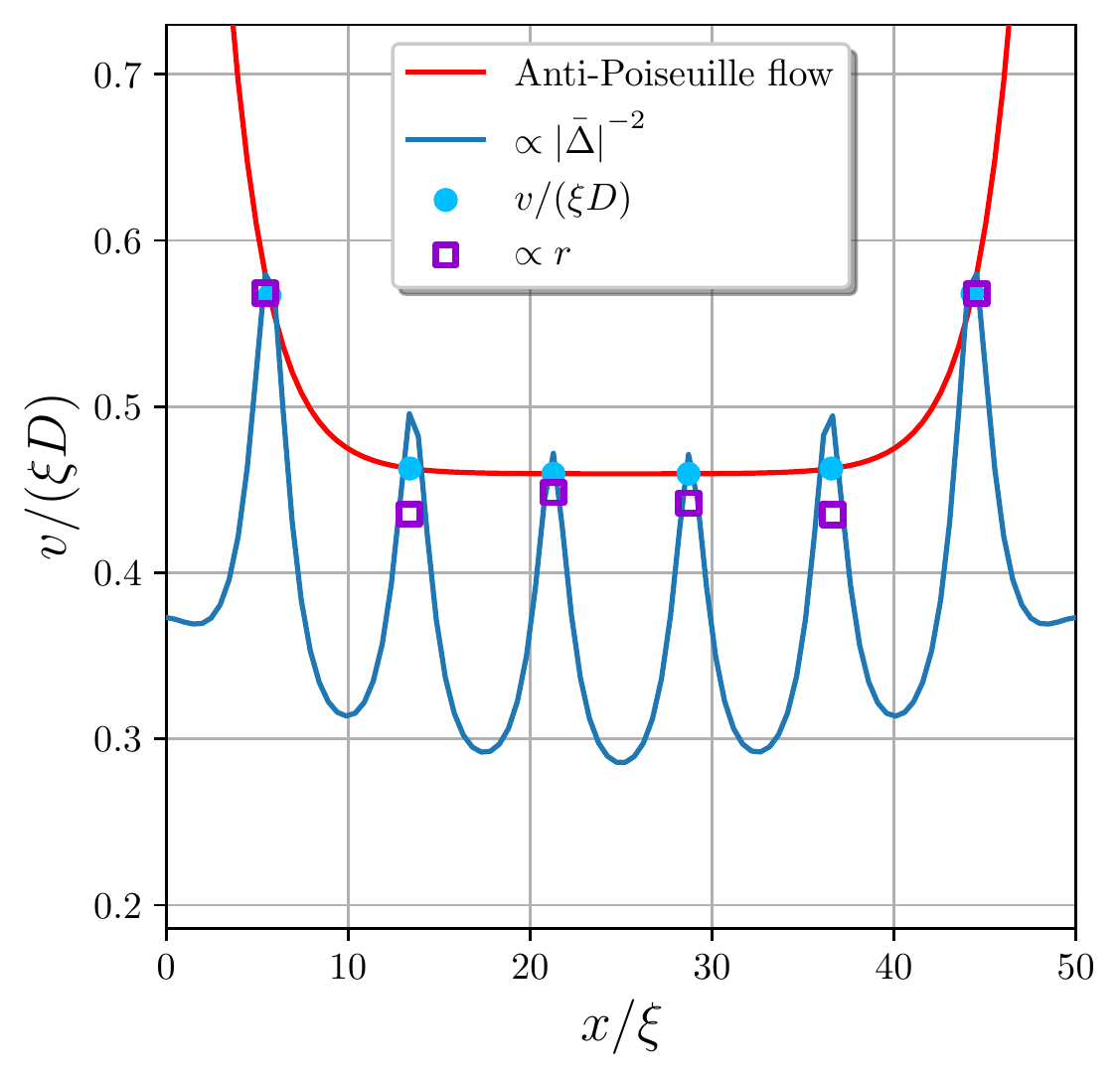}
\caption{Average speed of the vortex flow in dependence of their spatial position along $x$. The same averaging procedures are used for the vortex speed $v$ and the $x-$dependent order parameter $\bar{|\Delta|}$ as Fig. \ref{fig:vel1} and Fig. \ref{fig:gapvsJ}, respectively. To obtain a quantification of Anti-Poiseuille flow, we fit the vortex speeds by $C_0 + C_1 \cosh( C_2(x-L/2)/\xi)$, yielding $C_0=0.456$, $C_1=3.32 \times 10^{-5}$, and $C_2=0.455$. The resistance over the separate channels $r=\Delta \bar \Theta$ is calculated by taking differences of the averaged scalar potential $\bar \Theta$ between each vortex channel (see Fig.~\ref{fig:theta} for detail), where $\bar{\Theta}$ is obtained from the same averaging procedure as for $\bar{|\Delta|}$. The same parameters are chosen for the simulation as in Fig.~\ref{fig:snap}.}
\label{fig:vel}
\end{center}
\end{figure}

In order to check the relation Eq. \eqref{eq:predv}, we summarize representative results in Fig.~\ref{fig:vel}, where we plot time averaged vortex speeds $v$ (for an analysis of the time dependent velocity $\dot{y}$, see section \ref{section:stick_slip_motion}) and $x$-dependent inverse of the square of the averaged order parameter $\bar{|\Delta|}^{-2}$. Overall, we find good agreement between the simple prediction Eq.~\eqref{eq:predv} and the full numerical simulations for $v$. The slight disagreement between $v/(\xi D)$ (blue round dot) and $\bar{|\Delta|}^{-2}$ (blue curve) may be due to the negligence of the intricate vortex-vortex interactions.
\begin{figure}[t]
\begin{center} 
\includegraphics[width=0.49\textwidth, angle=-0]{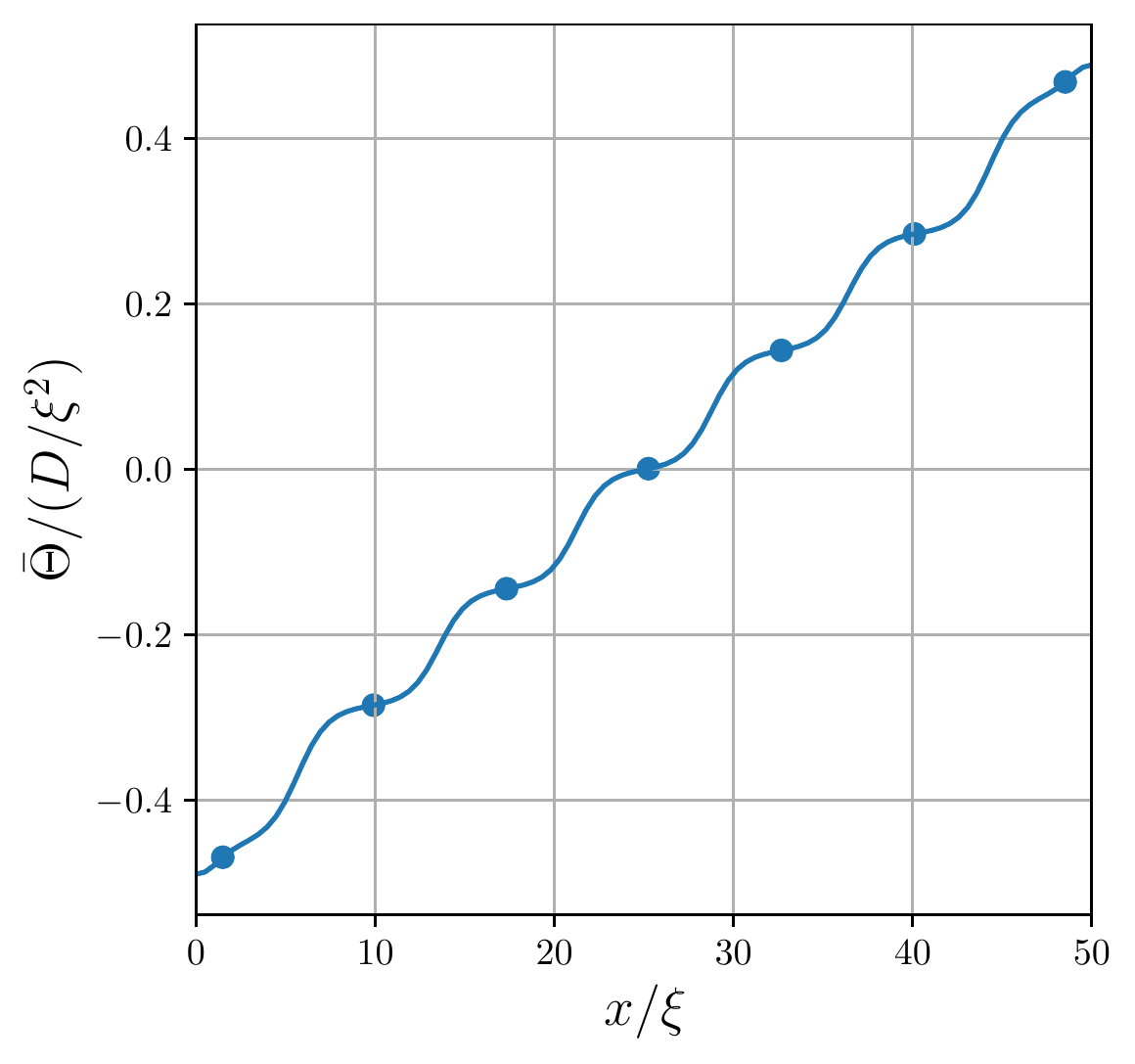}
\caption{Average scalar potential $\Bar{\Theta}$ in dependence of $x$. The same averaging procedure and parameters as in Fig.~\ref{fig:vel} are used. The quantity $r=\Delta \bar \Theta$ shown by the purple squares in Fig.~\ref{fig:vel} are calculated from differences between height of each plateau shown. Specifically, we use the points marked by the blue dots to define $r$.}
\label{fig:theta}
\end{center}
\end{figure}

The vortex flow behaves oppositely to the Poiseuille flow of water in a pipe. Instead of moving slower towards the boundary, the vortices tend to speed up. We therefore call this type of flow {\it Anti-Poiseuille flow}. The quantitative details of the flow behavior of course depend on the geometry and parameter regime considered. 
To quantify this effect, we note that in our simulations the behavior can be fitted by an exponential function $v=C_0 + C_1 \cosh( C_2\Delta x)$, where $\Delta x$ is the distance from the center of a sample and $C_{0/1/2}$ are constants. Specifically, we fit the vortex speeds by $C_0 + C_1 \cosh( C_2(x-L/2)/\xi)$. 
In this language $C_1<0$ would correspond to vortices moving slower approaching the boundary
, while $C_1>0$ means that vortices move faster at the edges (Anti-Poiseuille flow), with $|C_1|$ and $|C_2|$ quantifying the strength of the influence of the boundary.

\subsection{Resistivity}
To connect to experimentally accessible transport quantities we consider the resistances across the vortex channels $R=\Delta \bar \Theta/Q$, where $\Delta \bar \Theta$ is the difference of the scalar potential left and right to a given channel averaged over the y-direction.\cite{Machida1993} In a Bardeen-Stephen picture, that difference of the scalar potential is proportional to the number of vortices in each channel and their velocities. In Fig.~\ref{fig:theta}, we show the time and y-direction averaged scalar potential depending on $x$. Since the external current is constant in our system, we can consider differences of the scalar potential $r=\Delta \bar \Theta$ to be directly proportional to the resistance between each vortex channels. 
Fig.~\ref{fig:vel} clearly shows that faster vortex flows (blue dot) result in proportionally higher resistance 
$r$ (purple square) in accord with the Bardeen-Stephen picture. Following the same arguments as for the vortex velocity we therefore find  $r \propto \Delta^{-2}$. 

\subsection{Intermediate drive: stick-slip motion}
\label{section:stick_slip_motion}
\begin{figure}[t!]
\begin{center} 
\includegraphics[width=0.45\textwidth, angle=-0]{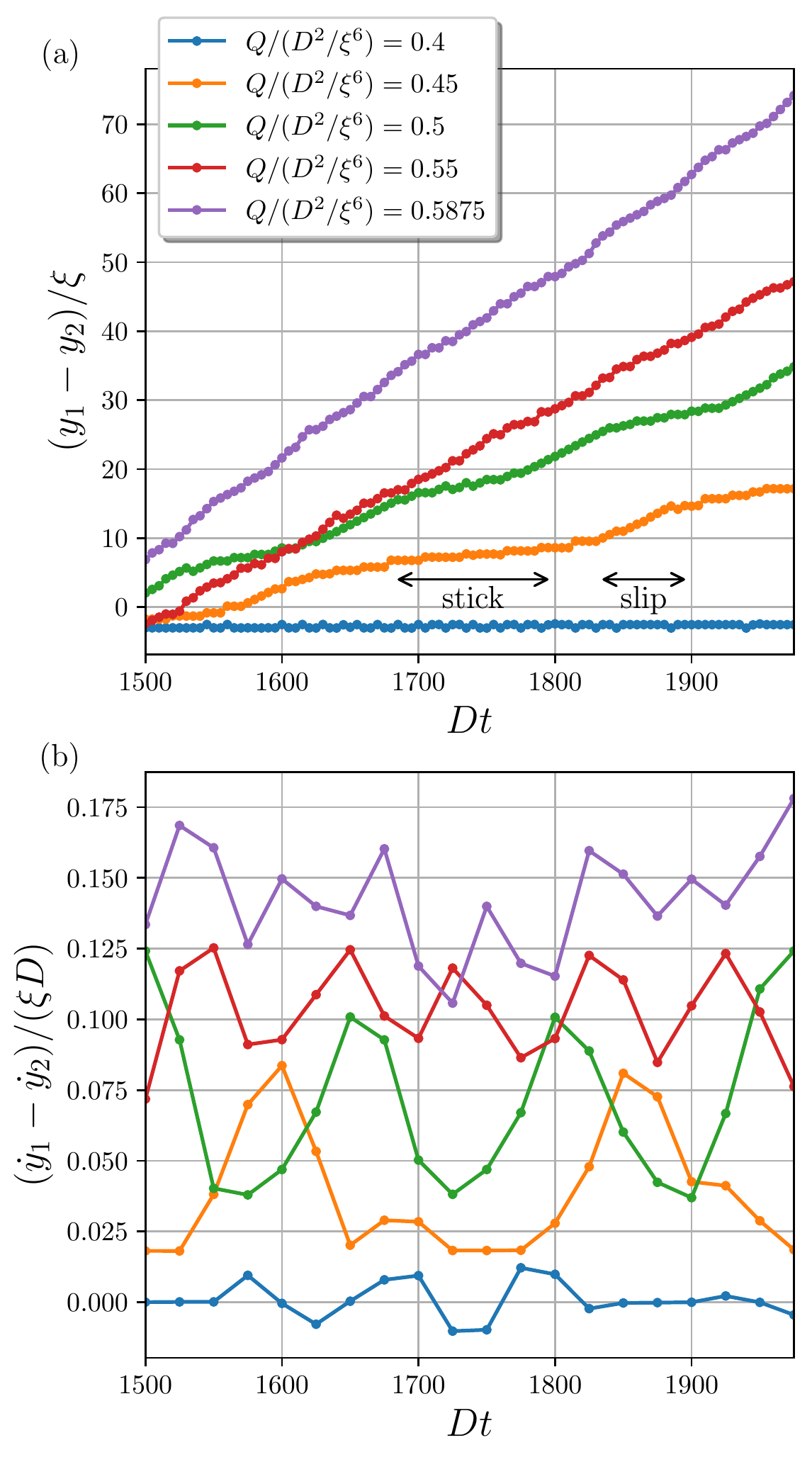}
\caption{(a) Differences of vortices' positions along the $y$-direction in the left most $y_{1}$ and the adjacent $y_{2}$ channel as a function of time (e.g. difference between $y$ coordinates of the vortex marked in red and blue in Fig.~\ref{fig:snap}). The periodic boundary condition is unfolded to calculate the $y$ coordinates. (b) Corresponding time derivative of (a), namely differences of the time dependent velocity within the left most $\dot{y}_{1}$ and the adjacent $\dot{y}_{2}$. All corresponding videos are available in the SM.\cite{SM} (see also TABLE \ref{table:video} in Appendix \ref{appendix:video}).
}
\label{fig:C1vsJ}
\end{center}
\end{figure}

Fig. \ref{fig:C1vsJ} shows how the difference between the position along the $y$-direction of the left most $y_{1}$ and the adjacent $y_{2}$ vortices evolves as a function of time $Dt$. For weaker drive ($Q/(D^2/\xi^6)=0.4$) the two vortices move together ($(y_1-y_2)/\xi$ is time independent). For the strongest drive ($Q/(D^2/\xi^6)=0.5875$) the two vortices move at different approximately constant velocities, but at intermediate drive a ``stick-slip" behavior is evident, in which the relative position exhibits a series of approximately linear increase in the difference of the positions of neighboring vortices separated by plateaus where the separation is time independent. 

In panel (b) we present the difference of the corresponding time derivatives. 
For the smallest $Q/(D^2/\xi^6)=0.4$, the difference in time dependent velocities $( \dot{y}_{1}- \dot{y}_{2})/ (\xi D)$ is almost zero at all times meaning that the lattice moves as a whole, while for the largest $Q/(D^2/\xi^6)\gtrsim 0.55$, the velocity difference $( \dot{y}_{1}- \dot{y}_{2})/ (\xi D)$ is essentially constant. For the intermediate $Q$ the relative velocity alternates in time between a large and a small value.


The times of linear increase in panel (a) with the peaks in panel (b) correspond to when the left most vortices passes the adjacent while the plateau in panel (a), (b) corresponds to the left most and the adjacent vortex moving with the same speeds as in the small sourced current regime $Q/(D^2/\xi^6)\leq0.4$ (see the SM video10, 12~\cite{SM}). At each region of the linear increase in panel (a) and the peak in (b), a shift of the left most and the adjacent vortices channels occurs. Or put into the language of the stick-slip phenomenology: initially the left most vortex sticks to the lattice, then it accelerates and passes (slips) across the adjacent vortex, and slows down to stick again to the newly established Abrikosov lattice, which is 
repeated as a converged non-equilibrium stationary state.
This behaviour relies on an inter-play between the vortex-vortex interactions and the large driving current (creating the reduced order parameter) as well as the boundary. The former tries to keep the Abrikosov vortex lattice while the latter two enforce the Anti-Poiseuille flow.
By increasing $Q/(D^2/\xi^6)$, the plateaus are disappearing in panel (a) and in (b) the number of peak is increased with the decrease of the peak height, which shows that vortex-vortex interaction is becoming gradually almost negligible and the speed difference between the left most and the adjacent vortices converges to a constant. So for increasing the strength of the source the stick-slip phenomenology is increasingly washed out.    

\section{Conclusion}
\label{conclusion}
Using time-dependent Ginzburg Landau theory, we study dynamics of vortex flows in 2D superconducting thin films with an external current sourced through the sample. We find that at small sourced currents the vortex lattice moves as a whole where all vortices move at the same velocity, while at larger sourced currents a vortex motion is assigned into channels where vortices in channels closer to the sample edge move faster than those farther way from the sample edge. We show that the velocity differential in the larger current regime arises from a position dependence of the viscosity depicted as by a Bardeen-Stephen picture and express the equation of motion for the vortex in the channel next to the boundary. 
Further analysis with neglecting intricate vortex-vortex interactions yields a simple prediction for the position dependent vortex velocity, which demonstrates a good agreement between this simple prediction and our full numerical calculations within the time-averaged analysis. 
The behaviour of vortex flow is found to be opposite from what is typical in Poiseuille-like flow, namely the velocity of the flow increases towards the boundary. In addition, detailed real-time analysis of the vortices' motions for the larger current regime reveals a stick-slip motion of the vortex lattice as the sourced current is increased. Initially, the vortices in the channel at the edges stick to the Abrikosov lattice, then they start to accelerate, move past their neighbours, followed by slowing down to again stick to another reestablished Abrikosov lattice, which is periodically continuing as a stationary state of the system. This intricate motion relies on the interplay between the vortex-vortex interactions and the large driving current at a boundary (creating the reduced order parameter).

Experimentally, this effect may be observed by a local probe that can count the rate at which vortices cross the probe to map the vortex velocity in space along the edge. A scanning superconducting quantum interference device (SQUID) can be used for low Ginzburg-Landau parameter superconductors,\cite{embon2017imaging} $\kappa=\Lambda/\xi$ (where $\Lambda$ and $\xi$ are the penetration depth and coherent length, respectively), or one could use an ultra-fast STM.\cite{khusnatdinov2000ultrafast} 

Although for all of the simulations in the present study, the insulator-superconductor boundary is considered as a choice of the boundary condition, there is no reason to believe that general features shown in this paper cannot be observed in real metal-superconductor boundary, which naturally creates the same scenario shown in this paper since the order parameter continuously goes to zero at the metal-superconductor boundary.\cite{work_progress}

\begin{acknowledgments} 
\noindent \textit{Acknowledgments}.---
This work was supported by the Deutsche
Forschungsgemeinschaft (DFG, German Research Foundation) via RTG 1995 and Germany’s Excellence Strategy - Cluster of Excellence Matter and Light for Quantum Computing (ML4Q) EXC 2004/1 - 390534769. AB and AJM acknowledge support from the NSF MRSEC program through the Center for Precision Assembly of Superstratic and Superatomic Solids (DMR-1420634). Simulations were performed with computing resources granted by RWTH Aachen University under project rwth0601 and rwth0507.  The Flatiron Institute is a division of the Simons Foundation. 

\end{acknowledgments}

\appendix
\section{Short explanation of the videos}
\label{appendix:video}
\begin{table}[t!]
\begin{center}
\caption{Correspondence between the video number and the parameter $Q$ in the SM~\cite{SM}}
{
\begin{tabular}{ccccccc}\hline\hline
video  &&  quantity & $Q/(D^2/\xi^6)$ \\ \hline
1    && $|\Delta|$ & 0.05  \\
2    && $|\Delta|$ & 0.10  \\
3    && $|\Delta|$ & 0.15  \\
4    && $|\Delta|$ & 0.20  \\
5    && $|\Delta|$ & 0.25  \\
6    && $|\Delta|$ & 0.30  \\
7    && $|\Delta|$ & 0.35  \\
8    && $|\Delta|$ & 0.40  \\
9    && $|\Delta|$ & 0.425  \\
10    && $|\Delta|$ & 0.45  \\
11    && $|\Delta|$ & 0.475  \\
12    && $|\Delta|$ & 0.5  \\
13    && $|\Delta|$ & 0.525  \\
14    && $|\Delta|$ & 0.55  \\
15    && $|\Delta|$ & 0.575  \\
16    && $|\Delta|$ & 0.5875  \\
17    && $|\Delta|$ & 0.6  \\
18    && $|\Delta|$ & 0.625  \\
19    && $|\Delta|$ & 0.65  \\\hline
\end{tabular}
}
\label{table:video}
\end{center}
\end{table}
In this section, we show the correspondence between the video number and a strength of the source term $Q$ in the SM. This is summarized in TABLE \ref{table:video}.

\bibliography{reference}%

\providecommand{\noopsort}[1]{}\providecommand{\singleletter}[1]{#1}%
\begin{thebibliography}{44}%
\makeatletter
\providecommand \@ifxundefined [1]{%
 \@ifx{#1\undefined}
}%
\providecommand \@ifnum [1]{%
 \ifnum #1\expandafter \@firstoftwo
 \else \expandafter \@secondoftwo
 \fi
}%
\providecommand \@ifx [1]{%
 \ifx #1\expandafter \@firstoftwo
 \else \expandafter \@secondoftwo
 \fi
}%
\providecommand \natexlab [1]{#1}%
\providecommand \enquote  [1]{``#1''}%
\providecommand \bibnamefont  [1]{#1}%
\providecommand \bibfnamefont [1]{#1}%
\providecommand \citenamefont [1]{#1}%
\providecommand \href@noop [0]{\@secondoftwo}%
\providecommand \href [0]{\begingroup \@sanitize@url \@href}%
\providecommand \@href[1]{\@@startlink{#1}\@@href}%
\providecommand \@@href[1]{\endgroup#1\@@endlink}%
\providecommand \@sanitize@url [0]{\catcode `\\12\catcode `\$12\catcode
  `\&12\catcode `\#12\catcode `\^12\catcode `\_12\catcode `\%12\relax}%
\providecommand \@@startlink[1]{}%
\providecommand \@@endlink[0]{}%
\providecommand \url  [0]{\begingroup\@sanitize@url \@url }%
\providecommand \@url [1]{\endgroup\@href {#1}{\urlprefix }}%
\providecommand \urlprefix  [0]{URL }%
\providecommand \Eprint [0]{\href }%
\providecommand \doibase [0]{http://dx.doi.org/}%
\providecommand \selectlanguage [0]{\@gobble}%
\providecommand \bibinfo  [0]{\@secondoftwo}%
\providecommand \bibfield  [0]{\@secondoftwo}%
\providecommand \translation [1]{[#1]}%
\providecommand \BibitemOpen [0]{}%
\providecommand \bibitemStop [0]{}%
\providecommand \bibitemNoStop [0]{.\EOS\space}%
\providecommand \EOS [0]{\spacefactor3000\relax}%
\providecommand \BibitemShut  [1]{\csname bibitem#1\endcsname}%
\let\auto@bib@innerbib\@empty
\bibitem [{\citenamefont {Abrikosov}(1957)}]{Abrikosov1957}%
  \BibitemOpen
  \bibfield  {author} {\bibinfo {author} {\bibfnamefont {A.}~\bibnamefont
  {Abrikosov}},\ }\href {\doibase https://doi.org/10.1016/0022-3697(57)90083-5}
  {\bibfield  {journal} {\bibinfo  {journal} {Journal of Physics and Chemistry
  of Solids}\ }\textbf {\bibinfo {volume} {2}},\ \bibinfo {pages} {199 }
  (\bibinfo {year} {1957})}\BibitemShut {NoStop}%
\bibitem [{\citenamefont {Bardeen}\ and\ \citenamefont
  {Stephen}(1965)}]{Bardeen1965}%
  \BibitemOpen
  \bibfield  {author} {\bibinfo {author} {\bibfnamefont {J.}~\bibnamefont
  {Bardeen}}\ and\ \bibinfo {author} {\bibfnamefont {M.~J.}\ \bibnamefont
  {Stephen}},\ }\href {\doibase 10.1103/PhysRev.140.A1197} {\bibfield
  {journal} {\bibinfo  {journal} {Phys. Rev.}\ }\textbf {\bibinfo {volume}
  {140}},\ \bibinfo {pages} {A1197} (\bibinfo {year} {1965})}\BibitemShut
  {NoStop}%
\bibitem [{\citenamefont {Tinkham}(2004)}]{Tinkham}%
  \BibitemOpen
  \bibfield  {author} {\bibinfo {author} {\bibfnamefont {M.}~\bibnamefont
  {Tinkham}},\ }\href {https://books.google.de/books?id=VpUk3NfwDIkC} {\emph
  {\bibinfo {title} {Introduction to Superconductivity}}},\ Dover Books on
  Physics Series\ (\bibinfo  {publisher} {Dover Publications},\ \bibinfo {year}
  {2004})\BibitemShut {NoStop}%
\bibitem [{\citenamefont {Devoret}\ and\ \citenamefont
  {Schoelkopf}(2013)}]{Devoret_2013}%
  \BibitemOpen
  \bibfield  {author} {\bibinfo {author} {\bibfnamefont {M.~H.}\ \bibnamefont
  {Devoret}}\ and\ \bibinfo {author} {\bibfnamefont {R.~J.}\ \bibnamefont
  {Schoelkopf}},\ }\href {\doibase 10.1126/science.1231930} {\bibfield
  {journal} {\bibinfo  {journal} {Science}\ }\textbf {\bibinfo {volume}
  {339}},\ \bibinfo {pages} {1169} (\bibinfo {year} {2013})}\BibitemShut
  {NoStop}%
\bibitem [{\citenamefont {Shaw}\ \emph {et~al.}(2019)\citenamefont {Shaw},
  \citenamefont {Blanco~Alvarez}, \citenamefont {Brisbois}, \citenamefont
  {Burger}, \citenamefont {Pinheiro}, \citenamefont {Kramer}, \citenamefont
  {Motta}, \citenamefont {Fleury-Frenette}, \citenamefont {Ortiz},
  \citenamefont {Vanderheyden},\ and\ \citenamefont {Silhanek}}]{met9101022}%
  \BibitemOpen
  \bibfield  {author} {\bibinfo {author} {\bibfnamefont {G.}~\bibnamefont
  {Shaw}}, \bibinfo {author} {\bibfnamefont {S.}~\bibnamefont
  {Blanco~Alvarez}}, \bibinfo {author} {\bibfnamefont {J.}~\bibnamefont
  {Brisbois}}, \bibinfo {author} {\bibfnamefont {L.}~\bibnamefont {Burger}},
  \bibinfo {author} {\bibfnamefont {L.~B. L.~G.}\ \bibnamefont {Pinheiro}},
  \bibinfo {author} {\bibfnamefont {R.~B.~G.}\ \bibnamefont {Kramer}}, \bibinfo
  {author} {\bibfnamefont {M.}~\bibnamefont {Motta}}, \bibinfo {author}
  {\bibfnamefont {K.}~\bibnamefont {Fleury-Frenette}}, \bibinfo {author}
  {\bibfnamefont {W.~A.}\ \bibnamefont {Ortiz}}, \bibinfo {author}
  {\bibfnamefont {B.}~\bibnamefont {Vanderheyden}}, \ and\ \bibinfo {author}
  {\bibfnamefont {A.~V.}\ \bibnamefont {Silhanek}},\ }\href {\doibase
  10.3390/met9101022} {\bibfield  {journal} {\bibinfo  {journal} {Metals}\
  }\textbf {\bibinfo {volume} {9}} (\bibinfo {year} {2019}),\
  10.3390/met9101022}\BibitemShut {NoStop}%
\bibitem [{\citenamefont {Si}\ \emph {et~al.}(2013)\citenamefont {Si},
  \citenamefont {Han}, \citenamefont {Shi}, \citenamefont {Ehrlich},
  \citenamefont {Jaroszynski}, \citenamefont {Goyal},\ and\ \citenamefont
  {Li}}]{Si_2013}%
  \BibitemOpen
  \bibfield  {author} {\bibinfo {author} {\bibfnamefont {W.}~\bibnamefont
  {Si}}, \bibinfo {author} {\bibfnamefont {S.~J.}\ \bibnamefont {Han}},
  \bibinfo {author} {\bibfnamefont {X.}~\bibnamefont {Shi}}, \bibinfo {author}
  {\bibfnamefont {S.~N.}\ \bibnamefont {Ehrlich}}, \bibinfo {author}
  {\bibfnamefont {J.}~\bibnamefont {Jaroszynski}}, \bibinfo {author}
  {\bibfnamefont {A.}~\bibnamefont {Goyal}}, \ and\ \bibinfo {author}
  {\bibfnamefont {Q.}~\bibnamefont {Li}},\ }\href {\doibase 10.1038/ncomms2337}
  {\bibfield  {journal} {\bibinfo  {journal} {Nature Communications}\ }\textbf
  {\bibinfo {volume} {4}} (\bibinfo {year} {2013}),\
  10.1038/ncomms2337}\BibitemShut {NoStop}%
\bibitem [{\citenamefont {Welp}\ \emph {et~al.}(2013)\citenamefont {Welp},
  \citenamefont {Kadowaki},\ and\ \citenamefont {Kleiner}}]{Welp_2013}%
  \BibitemOpen
  \bibfield  {author} {\bibinfo {author} {\bibfnamefont {U.}~\bibnamefont
  {Welp}}, \bibinfo {author} {\bibfnamefont {K.}~\bibnamefont {Kadowaki}}, \
  and\ \bibinfo {author} {\bibfnamefont {R.}~\bibnamefont {Kleiner}},\ }\href
  {\doibase 10.1038/nphoton.2013.216} {\bibfield  {journal} {\bibinfo
  {journal} {Nature Photonics}\ }\textbf {\bibinfo {volume} {7}},\ \bibinfo
  {pages} {702} (\bibinfo {year} {2013})}\BibitemShut {NoStop}%
\bibitem [{\citenamefont {Padamsee}(2001)}]{Padamsee_2001}%
  \BibitemOpen
  \bibfield  {author} {\bibinfo {author} {\bibfnamefont {H.}~\bibnamefont
  {Padamsee}},\ }\href {\doibase 10.1088/0953-2048/14/4/202} {\bibfield
  {journal} {\bibinfo  {journal} {Superconductor Science and Technology}\
  }\textbf {\bibinfo {volume} {14}},\ \bibinfo {pages} {R28} (\bibinfo {year}
  {2001})}\BibitemShut {NoStop}%
\bibitem [{\citenamefont {Gurevich}\ and\ \citenamefont
  {Ciovati}(2008)}]{Gurevich2008}%
  \BibitemOpen
  \bibfield  {author} {\bibinfo {author} {\bibfnamefont {A.}~\bibnamefont
  {Gurevich}}\ and\ \bibinfo {author} {\bibfnamefont {G.}~\bibnamefont
  {Ciovati}},\ }\href {\doibase 10.1103/PhysRevB.77.104501} {\bibfield
  {journal} {\bibinfo  {journal} {Phys. Rev. B}\ }\textbf {\bibinfo {volume}
  {77}},\ \bibinfo {pages} {104501} (\bibinfo {year} {2008})}\BibitemShut
  {NoStop}%
\bibitem [{\citenamefont {Veshchunov}\ \emph {et~al.}(2016)\citenamefont
  {Veshchunov}, \citenamefont {Magrini}, \citenamefont {Mironov}, \citenamefont
  {Godin}, \citenamefont {Trebbia}, \citenamefont {Buzdin}, \citenamefont
  {Tamarat},\ and\ \citenamefont {Lounis}}]{Veshchunov_2016}%
  \BibitemOpen
  \bibfield  {author} {\bibinfo {author} {\bibfnamefont {I.~S.}\ \bibnamefont
  {Veshchunov}}, \bibinfo {author} {\bibfnamefont {W.}~\bibnamefont {Magrini}},
  \bibinfo {author} {\bibfnamefont {S.~V.}\ \bibnamefont {Mironov}}, \bibinfo
  {author} {\bibfnamefont {A.~G.}\ \bibnamefont {Godin}}, \bibinfo {author}
  {\bibfnamefont {J.-B.}\ \bibnamefont {Trebbia}}, \bibinfo {author}
  {\bibfnamefont {A.~I.}\ \bibnamefont {Buzdin}}, \bibinfo {author}
  {\bibfnamefont {P.}~\bibnamefont {Tamarat}}, \ and\ \bibinfo {author}
  {\bibfnamefont {B.}~\bibnamefont {Lounis}},\ }\href {\doibase
  10.1038/ncomms12801} {\bibfield  {journal} {\bibinfo  {journal} {Nature
  Communications}\ }\textbf {\bibinfo {volume} {7}} (\bibinfo {year} {2016}),\
  10.1038/ncomms12801}\BibitemShut {NoStop}%
\bibitem [{\citenamefont {Madan}\ \emph {et~al.}(2018)\citenamefont {Madan},
  \citenamefont {Buh}, \citenamefont {Baranov}, \citenamefont {Kabanov},
  \citenamefont {Mrzel},\ and\ \citenamefont {Mihailovic}}]{Madan_2018}%
  \BibitemOpen
  \bibfield  {author} {\bibinfo {author} {\bibfnamefont {I.}~\bibnamefont
  {Madan}}, \bibinfo {author} {\bibfnamefont {J.}~\bibnamefont {Buh}}, \bibinfo
  {author} {\bibfnamefont {V.~V.}\ \bibnamefont {Baranov}}, \bibinfo {author}
  {\bibfnamefont {V.~V.}\ \bibnamefont {Kabanov}}, \bibinfo {author}
  {\bibfnamefont {A.}~\bibnamefont {Mrzel}}, \ and\ \bibinfo {author}
  {\bibfnamefont {D.}~\bibnamefont {Mihailovic}},\ }\href {\doibase
  10.1126/sciadv.aao0043} {\bibfield  {journal} {\bibinfo  {journal} {Science
  Advances}\ }\textbf {\bibinfo {volume} {4}},\ \bibinfo {pages} {eaao0043}
  (\bibinfo {year} {2018})}\BibitemShut {NoStop}%
\bibitem [{\citenamefont {Rouco}\ \emph {et~al.}(2019)\citenamefont {Rouco},
  \citenamefont {Navau}, \citenamefont {Del-Valle}, \citenamefont {Massarotti},
  \citenamefont {Papari}, \citenamefont {Stornaiuolo}, \citenamefont
  {Obradors}, \citenamefont {Puig}, \citenamefont {Tafuri}, \citenamefont
  {Sanchez},\ and\ \citenamefont {Palau}}]{Rouco2019}%
  \BibitemOpen
  \bibfield  {author} {\bibinfo {author} {\bibfnamefont {V.}~\bibnamefont
  {Rouco}}, \bibinfo {author} {\bibfnamefont {C.}~\bibnamefont {Navau}},
  \bibinfo {author} {\bibfnamefont {N.}~\bibnamefont {Del-Valle}}, \bibinfo
  {author} {\bibfnamefont {D.}~\bibnamefont {Massarotti}}, \bibinfo {author}
  {\bibfnamefont {G.~P.}\ \bibnamefont {Papari}}, \bibinfo {author}
  {\bibfnamefont {D.}~\bibnamefont {Stornaiuolo}}, \bibinfo {author}
  {\bibfnamefont {X.}~\bibnamefont {Obradors}}, \bibinfo {author}
  {\bibfnamefont {T.}~\bibnamefont {Puig}}, \bibinfo {author} {\bibfnamefont
  {F.}~\bibnamefont {Tafuri}}, \bibinfo {author} {\bibfnamefont
  {A.}~\bibnamefont {Sanchez}}, \ and\ \bibinfo {author} {\bibfnamefont
  {A.}~\bibnamefont {Palau}},\ }\href {\doibase 10.1021/acs.nanolett.9b01693}
  {\bibfield  {journal} {\bibinfo  {journal} {Nano Letters}\ }\textbf {\bibinfo
  {volume} {19}},\ \bibinfo {pages} {4174} (\bibinfo {year} {2019})},\ \Eprint
  {http://arxiv.org/abs/https://doi.org/10.1021/acs.nanolett.9b01693}
  {https://doi.org/10.1021/acs.nanolett.9b01693} \BibitemShut {NoStop}%
\bibitem [{\citenamefont {Tamir}\ \emph {et~al.}(2019)\citenamefont {Tamir},
  \citenamefont {Benyamini}, \citenamefont {Telford}, \citenamefont
  {Gorniaczyk}, \citenamefont {Doron}, \citenamefont {Levinson}, \citenamefont
  {Wang}, \citenamefont {Gay}, \citenamefont {Sac{\'e}p{\'e}}, \citenamefont
  {Hone}, \citenamefont {Watanabe}, \citenamefont {Taniguchi}, \citenamefont
  {Dean}, \citenamefont {Pasupathy},\ and\ \citenamefont
  {Shahar}}]{Tamireaau2019}%
  \BibitemOpen
  \bibfield  {author} {\bibinfo {author} {\bibfnamefont {I.}~\bibnamefont
  {Tamir}}, \bibinfo {author} {\bibfnamefont {A.}~\bibnamefont {Benyamini}},
  \bibinfo {author} {\bibfnamefont {E.~J.}\ \bibnamefont {Telford}}, \bibinfo
  {author} {\bibfnamefont {F.}~\bibnamefont {Gorniaczyk}}, \bibinfo {author}
  {\bibfnamefont {A.}~\bibnamefont {Doron}}, \bibinfo {author} {\bibfnamefont
  {T.}~\bibnamefont {Levinson}}, \bibinfo {author} {\bibfnamefont
  {D.}~\bibnamefont {Wang}}, \bibinfo {author} {\bibfnamefont {F.}~\bibnamefont
  {Gay}}, \bibinfo {author} {\bibfnamefont {B.}~\bibnamefont {Sac{\'e}p{\'e}}},
  \bibinfo {author} {\bibfnamefont {J.}~\bibnamefont {Hone}}, \bibinfo {author}
  {\bibfnamefont {K.}~\bibnamefont {Watanabe}}, \bibinfo {author}
  {\bibfnamefont {T.}~\bibnamefont {Taniguchi}}, \bibinfo {author}
  {\bibfnamefont {C.~R.}\ \bibnamefont {Dean}}, \bibinfo {author}
  {\bibfnamefont {A.~N.}\ \bibnamefont {Pasupathy}}, \ and\ \bibinfo {author}
  {\bibfnamefont {D.}~\bibnamefont {Shahar}},\ }\href {\doibase
  10.1126/sciadv.aau3826} {\bibfield  {journal} {\bibinfo  {journal} {Science
  Advances}\ }\textbf {\bibinfo {volume} {5}} (\bibinfo {year} {2019}),\
  10.1126/sciadv.aau3826},\ \Eprint
  {http://arxiv.org/abs/https://advances.sciencemag.org/content/5/3/eaau3826.full.pdf}
  {https://advances.sciencemag.org/content/5/3/eaau3826.full.pdf} \BibitemShut
  {NoStop}%
\bibitem [{\citenamefont {Embon}\ \emph
  {et~al.}(2017{\natexlab{a}})\citenamefont {Embon}, \citenamefont {Anahory},
  \citenamefont {Jeli{\'{c}}}, \citenamefont {Lachman}, \citenamefont
  {Myasoedov}, \citenamefont {Huber}, \citenamefont {Mikitik}, \citenamefont
  {Silhanek}, \citenamefont {Milo{\v{s}}evi{\'{c}}}, \citenamefont {Gurevich},\
  and\ \citenamefont {Zeldov}}]{Embon2017}%
  \BibitemOpen
  \bibfield  {author} {\bibinfo {author} {\bibfnamefont {L.}~\bibnamefont
  {Embon}}, \bibinfo {author} {\bibfnamefont {Y.}~\bibnamefont {Anahory}},
  \bibinfo {author} {\bibfnamefont {{\v{Z}}.}~\bibnamefont {Jeli{\'{c}}}},
  \bibinfo {author} {\bibfnamefont {E.~O.}\ \bibnamefont {Lachman}}, \bibinfo
  {author} {\bibfnamefont {Y.}~\bibnamefont {Myasoedov}}, \bibinfo {author}
  {\bibfnamefont {M.~E.}\ \bibnamefont {Huber}}, \bibinfo {author}
  {\bibfnamefont {G.~P.}\ \bibnamefont {Mikitik}}, \bibinfo {author}
  {\bibfnamefont {A.~V.}\ \bibnamefont {Silhanek}}, \bibinfo {author}
  {\bibfnamefont {M.~V.}\ \bibnamefont {Milo{\v{s}}evi{\'{c}}}}, \bibinfo
  {author} {\bibfnamefont {A.}~\bibnamefont {Gurevich}}, \ and\ \bibinfo
  {author} {\bibfnamefont {E.}~\bibnamefont {Zeldov}},\ }\href {\doibase
  10.1038/s41467-017-00089-3} {\bibfield  {journal} {\bibinfo  {journal}
  {Nature Communications}\ }\textbf {\bibinfo {volume} {8}} (\bibinfo {year}
  {2017}{\natexlab{a}}),\ 10.1038/s41467-017-00089-3}\BibitemShut {NoStop}%
\bibitem [{\citenamefont {Dobrovolskiy}\ \emph {et~al.}(2020)\citenamefont
  {Dobrovolskiy}, \citenamefont {Vodolazov}, \citenamefont {Porrati},
  \citenamefont {Sachser}, \citenamefont {Bevz}, \citenamefont {Mikhailov},
  \citenamefont {Chumak},\ and\ \citenamefont {Huth}}]{Dobrovolskiy_2020}%
  \BibitemOpen
  \bibfield  {author} {\bibinfo {author} {\bibfnamefont {O.~V.}\ \bibnamefont
  {Dobrovolskiy}}, \bibinfo {author} {\bibfnamefont {D.~Y.}\ \bibnamefont
  {Vodolazov}}, \bibinfo {author} {\bibfnamefont {F.}~\bibnamefont {Porrati}},
  \bibinfo {author} {\bibfnamefont {R.}~\bibnamefont {Sachser}}, \bibinfo
  {author} {\bibfnamefont {V.~M.}\ \bibnamefont {Bevz}}, \bibinfo {author}
  {\bibfnamefont {M.~Y.}\ \bibnamefont {Mikhailov}}, \bibinfo {author}
  {\bibfnamefont {A.~V.}\ \bibnamefont {Chumak}}, \ and\ \bibinfo {author}
  {\bibfnamefont {M.}~\bibnamefont {Huth}},\ }\href {\doibase
  10.1038/s41467-020-16987-y} {\bibfield  {journal} {\bibinfo  {journal}
  {Nature Communications}\ }\textbf {\bibinfo {volume} {11}} (\bibinfo {year}
  {2020}),\ 10.1038/s41467-020-16987-y}\BibitemShut {NoStop}%
\bibitem [{\citenamefont {Vodolazov}\ and\ \citenamefont
  {Peeters}(2007)}]{Vodolazov2007}%
  \BibitemOpen
  \bibfield  {author} {\bibinfo {author} {\bibfnamefont {D.~Y.}\ \bibnamefont
  {Vodolazov}}\ and\ \bibinfo {author} {\bibfnamefont {F.~M.}\ \bibnamefont
  {Peeters}},\ }\href {\doibase 10.1103/PhysRevB.76.014521} {\bibfield
  {journal} {\bibinfo  {journal} {Phys. Rev. B}\ }\textbf {\bibinfo {volume}
  {76}},\ \bibinfo {pages} {014521} (\bibinfo {year} {2007})}\BibitemShut
  {NoStop}%
\bibitem [{\citenamefont {Vodolazov}(2013)}]{Vodolazov2013}%
  \BibitemOpen
  \bibfield  {author} {\bibinfo {author} {\bibfnamefont {D.~Y.}\ \bibnamefont
  {Vodolazov}},\ }\href {\doibase 10.1103/PhysRevB.88.014525} {\bibfield
  {journal} {\bibinfo  {journal} {Phys. Rev. B}\ }\textbf {\bibinfo {volume}
  {88}},\ \bibinfo {pages} {014525} (\bibinfo {year} {2013})}\BibitemShut
  {NoStop}%
\bibitem [{\citenamefont {Grimaldi}\ \emph {et~al.}(2015)\citenamefont
  {Grimaldi}, \citenamefont {Leo}, \citenamefont {Sabatino}, \citenamefont
  {Carapella}, \citenamefont {Nigro}, \citenamefont {Pace}, \citenamefont
  {Moshchalkov},\ and\ \citenamefont {Silhanek}}]{Grimaldi2015}%
  \BibitemOpen
  \bibfield  {author} {\bibinfo {author} {\bibfnamefont {G.}~\bibnamefont
  {Grimaldi}}, \bibinfo {author} {\bibfnamefont {A.}~\bibnamefont {Leo}},
  \bibinfo {author} {\bibfnamefont {P.}~\bibnamefont {Sabatino}}, \bibinfo
  {author} {\bibfnamefont {G.}~\bibnamefont {Carapella}}, \bibinfo {author}
  {\bibfnamefont {A.}~\bibnamefont {Nigro}}, \bibinfo {author} {\bibfnamefont
  {S.}~\bibnamefont {Pace}}, \bibinfo {author} {\bibfnamefont {V.~V.}\
  \bibnamefont {Moshchalkov}}, \ and\ \bibinfo {author} {\bibfnamefont {A.~V.}\
  \bibnamefont {Silhanek}},\ }\href {\doibase 10.1103/PhysRevB.92.024513}
  {\bibfield  {journal} {\bibinfo  {journal} {Phys. Rev. B}\ }\textbf {\bibinfo
  {volume} {92}},\ \bibinfo {pages} {024513} (\bibinfo {year}
  {2015})}\BibitemShut {NoStop}%
\bibitem [{\citenamefont {Berdiyorov}\ \emph {et~al.}(2014)\citenamefont
  {Berdiyorov}, \citenamefont {Harrabi}, \citenamefont {Oktasendra},
  \citenamefont {Gasmi}, \citenamefont {Mansour}, \citenamefont {Maneval},\
  and\ \citenamefont {Peeters}}]{Berdiyorov2014}%
  \BibitemOpen
  \bibfield  {author} {\bibinfo {author} {\bibfnamefont {G.}~\bibnamefont
  {Berdiyorov}}, \bibinfo {author} {\bibfnamefont {K.}~\bibnamefont {Harrabi}},
  \bibinfo {author} {\bibfnamefont {F.}~\bibnamefont {Oktasendra}}, \bibinfo
  {author} {\bibfnamefont {K.}~\bibnamefont {Gasmi}}, \bibinfo {author}
  {\bibfnamefont {A.~I.}\ \bibnamefont {Mansour}}, \bibinfo {author}
  {\bibfnamefont {J.~P.}\ \bibnamefont {Maneval}}, \ and\ \bibinfo {author}
  {\bibfnamefont {F.~M.}\ \bibnamefont {Peeters}},\ }\href {\doibase
  10.1103/PhysRevB.90.054506} {\bibfield  {journal} {\bibinfo  {journal} {Phys.
  Rev. B}\ }\textbf {\bibinfo {volume} {90}},\ \bibinfo {pages} {054506}
  (\bibinfo {year} {2014})}\BibitemShut {NoStop}%
\bibitem [{\citenamefont {Andronov}\ \emph {et~al.}(1993)\citenamefont
  {Andronov}, \citenamefont {Gordion}, \citenamefont {Kurin}, \citenamefont
  {Nefedov},\ and\ \citenamefont {Shereshevsky}}]{Andronov1993}%
  \BibitemOpen
  \bibfield  {author} {\bibinfo {author} {\bibfnamefont {A.}~\bibnamefont
  {Andronov}}, \bibinfo {author} {\bibfnamefont {I.}~\bibnamefont {Gordion}},
  \bibinfo {author} {\bibfnamefont {V.}~\bibnamefont {Kurin}}, \bibinfo
  {author} {\bibfnamefont {I.}~\bibnamefont {Nefedov}}, \ and\ \bibinfo
  {author} {\bibfnamefont {I.}~\bibnamefont {Shereshevsky}},\ }\href {\doibase
  https://doi.org/10.1016/0921-4534(93)90777-N} {\bibfield  {journal} {\bibinfo
   {journal} {Physica C: Superconductivity and its Applications}\ }\textbf
  {\bibinfo {volume} {213}},\ \bibinfo {pages} {193 } (\bibinfo {year}
  {1993})}\BibitemShut {NoStop}%
\bibitem [{\citenamefont {Sivakov}\ \emph {et~al.}(2003)\citenamefont
  {Sivakov}, \citenamefont {Glukhov}, \citenamefont {Omelyanchouk},
  \citenamefont {Koval}, \citenamefont {M\"uller},\ and\ \citenamefont
  {Ustinov}}]{Sivakov2003}%
  \BibitemOpen
  \bibfield  {author} {\bibinfo {author} {\bibfnamefont {A.~G.}\ \bibnamefont
  {Sivakov}}, \bibinfo {author} {\bibfnamefont {A.~M.}\ \bibnamefont
  {Glukhov}}, \bibinfo {author} {\bibfnamefont {A.~N.}\ \bibnamefont
  {Omelyanchouk}}, \bibinfo {author} {\bibfnamefont {Y.}~\bibnamefont {Koval}},
  \bibinfo {author} {\bibfnamefont {P.}~\bibnamefont {M\"uller}}, \ and\
  \bibinfo {author} {\bibfnamefont {A.~V.}\ \bibnamefont {Ustinov}},\ }\href
  {\doibase 10.1103/PhysRevLett.91.267001} {\bibfield  {journal} {\bibinfo
  {journal} {Phys. Rev. Lett.}\ }\textbf {\bibinfo {volume} {91}},\ \bibinfo
  {pages} {267001} (\bibinfo {year} {2003})}\BibitemShut {NoStop}%
\bibitem [{\citenamefont {Keizer}\ \emph {et~al.}(2006)\citenamefont {Keizer},
  \citenamefont {Flokstra}, \citenamefont {Aarts},\ and\ \citenamefont
  {Klapwijk}}]{Keizer2006}%
  \BibitemOpen
  \bibfield  {author} {\bibinfo {author} {\bibfnamefont {R.~S.}\ \bibnamefont
  {Keizer}}, \bibinfo {author} {\bibfnamefont {M.~G.}\ \bibnamefont
  {Flokstra}}, \bibinfo {author} {\bibfnamefont {J.}~\bibnamefont {Aarts}}, \
  and\ \bibinfo {author} {\bibfnamefont {T.~M.}\ \bibnamefont {Klapwijk}},\
  }\href {\doibase 10.1103/PhysRevLett.96.147002} {\bibfield  {journal}
  {\bibinfo  {journal} {Phys. Rev. Lett.}\ }\textbf {\bibinfo {volume} {96}},\
  \bibinfo {pages} {147002} (\bibinfo {year} {2006})}\BibitemShut {NoStop}%
\bibitem [{\citenamefont {Berdiyorov}\ \emph {et~al.}(2009)\citenamefont
  {Berdiyorov}, \citenamefont {Milo\ifmmode \check{s}\else
  \v{s}\fi{}evi\ifmmode~\acute{c}\else \'{c}\fi{}},\ and\ \citenamefont
  {Peeters}}]{Berdiyorov2009}%
  \BibitemOpen
  \bibfield  {author} {\bibinfo {author} {\bibfnamefont {G.~R.}\ \bibnamefont
  {Berdiyorov}}, \bibinfo {author} {\bibfnamefont {M.~V.}\ \bibnamefont
  {Milo\ifmmode \check{s}\else \v{s}\fi{}evi\ifmmode~\acute{c}\else
  \'{c}\fi{}}}, \ and\ \bibinfo {author} {\bibfnamefont {F.~M.}\ \bibnamefont
  {Peeters}},\ }\href {\doibase 10.1103/PhysRevB.79.184506} {\bibfield
  {journal} {\bibinfo  {journal} {Phys. Rev. B}\ }\textbf {\bibinfo {volume}
  {79}},\ \bibinfo {pages} {184506} (\bibinfo {year} {2009})}\BibitemShut
  {NoStop}%
\bibitem [{\citenamefont {Silhanek}\ \emph {et~al.}(2010)\citenamefont
  {Silhanek}, \citenamefont {Milo\ifmmode \check{s}\else
  \v{s}\fi{}evi\ifmmode~\acute{c}\else \'{c}\fi{}}, \citenamefont {Kramer},
  \citenamefont {Berdiyorov}, \citenamefont {Van~de Vondel}, \citenamefont
  {Luccas}, \citenamefont {Puig}, \citenamefont {Peeters},\ and\ \citenamefont
  {Moshchalkov}}]{Silhanek2010}%
  \BibitemOpen
  \bibfield  {author} {\bibinfo {author} {\bibfnamefont {A.~V.}\ \bibnamefont
  {Silhanek}}, \bibinfo {author} {\bibfnamefont {M.~V.}\ \bibnamefont
  {Milo\ifmmode \check{s}\else \v{s}\fi{}evi\ifmmode~\acute{c}\else
  \'{c}\fi{}}}, \bibinfo {author} {\bibfnamefont {R.~B.~G.}\ \bibnamefont
  {Kramer}}, \bibinfo {author} {\bibfnamefont {G.~R.}\ \bibnamefont
  {Berdiyorov}}, \bibinfo {author} {\bibfnamefont {J.}~\bibnamefont {Van~de
  Vondel}}, \bibinfo {author} {\bibfnamefont {R.~F.}\ \bibnamefont {Luccas}},
  \bibinfo {author} {\bibfnamefont {T.}~\bibnamefont {Puig}}, \bibinfo {author}
  {\bibfnamefont {F.~M.}\ \bibnamefont {Peeters}}, \ and\ \bibinfo {author}
  {\bibfnamefont {V.~V.}\ \bibnamefont {Moshchalkov}},\ }\href {\doibase
  10.1103/PhysRevLett.104.017001} {\bibfield  {journal} {\bibinfo  {journal}
  {Phys. Rev. Lett.}\ }\textbf {\bibinfo {volume} {104}},\ \bibinfo {pages}
  {017001} (\bibinfo {year} {2010})}\BibitemShut {NoStop}%
\bibitem [{\citenamefont {Larkin}\ and\ \citenamefont
  {Ovchinnikov}(1975)}]{Larkin1975}%
  \BibitemOpen
  \bibfield  {author} {\bibinfo {author} {\bibfnamefont {A.~I.}\ \bibnamefont
  {Larkin}}\ and\ \bibinfo {author} {\bibfnamefont {Y.}~\bibnamefont
  {Ovchinnikov}},\ }\href@noop {} {\bibfield  {journal} {\bibinfo  {journal}
  {Sov. Phys.-JETP}\ }\textbf {\bibinfo {volume} {41}},\ \bibinfo {pages}
  {960–} (\bibinfo {year} {1975})}\BibitemShut {NoStop}%
\bibitem [{\citenamefont {Gor’kov}(1959)}]{Gorkov1959}%
  \BibitemOpen
  \bibfield  {author} {\bibinfo {author} {\bibfnamefont {L.~P.}\ \bibnamefont
  {Gor’kov}},\ }\href@noop {} {\bibfield  {journal} {\bibinfo  {journal}
  {Sov. Phys.-JETP}\ }\textbf {\bibinfo {volume} {9}},\ \bibinfo {pages} {1364}
  (\bibinfo {year} {1959})}\BibitemShut {NoStop}%
\bibitem [{\citenamefont {Schmid}(1966)}]{Schmid1966}%
  \BibitemOpen
  \bibfield  {author} {\bibinfo {author} {\bibfnamefont {A.}~\bibnamefont
  {Schmid}},\ }\href {\doibase 10.1007/bf02422669} {\bibfield  {journal}
  {\bibinfo  {journal} {Physik der Kondensierten Materie}\ }\textbf {\bibinfo
  {volume} {5}},\ \bibinfo {pages} {302} (\bibinfo {year} {1966})}\BibitemShut
  {NoStop}%
\bibitem [{\citenamefont {Cyrot}(1973)}]{Cyrot1973}%
  \BibitemOpen
  \bibfield  {author} {\bibinfo {author} {\bibfnamefont {M.}~\bibnamefont
  {Cyrot}},\ }\href {\doibase 10.1088/0034-4885/36/2/001} {\bibfield  {journal}
  {\bibinfo  {journal} {Reports on Progress in Physics}\ }\textbf {\bibinfo
  {volume} {36}},\ \bibinfo {pages} {103} (\bibinfo {year} {1973})}\BibitemShut
  {NoStop}%
\bibitem [{\citenamefont {Kennes}\ and\ \citenamefont {Millis}(2017)}]{Kennes}%
  \BibitemOpen
  \bibfield  {author} {\bibinfo {author} {\bibfnamefont {D.~M.}\ \bibnamefont
  {Kennes}}\ and\ \bibinfo {author} {\bibfnamefont {A.~J.}\ \bibnamefont
  {Millis}},\ }\href {\doibase 10.1103/PhysRevB.96.064507} {\bibfield
  {journal} {\bibinfo  {journal} {Phys. Rev. B}\ }\textbf {\bibinfo {volume}
  {96}},\ \bibinfo {pages} {064507} (\bibinfo {year} {2017})}\BibitemShut
  {NoStop}%
\bibitem [{\citenamefont {Benyamini}\ \emph {et~al.}(2019)\citenamefont
  {Benyamini}, \citenamefont {Telford}, \citenamefont {Kennes}, \citenamefont
  {Wang}, \citenamefont {Williams}, \citenamefont {Watanabe}, \citenamefont
  {Taniguchi}, \citenamefont {Shahar}, \citenamefont {Hone}, \citenamefont
  {Dean}, \citenamefont {Millis},\ and\ \citenamefont
  {Pasupathy}}]{Benyamini_2019}%
  \BibitemOpen
  \bibfield  {author} {\bibinfo {author} {\bibfnamefont {A.}~\bibnamefont
  {Benyamini}}, \bibinfo {author} {\bibfnamefont {E.~J.}\ \bibnamefont
  {Telford}}, \bibinfo {author} {\bibfnamefont {D.~M.}\ \bibnamefont {Kennes}},
  \bibinfo {author} {\bibfnamefont {D.}~\bibnamefont {Wang}}, \bibinfo {author}
  {\bibfnamefont {A.}~\bibnamefont {Williams}}, \bibinfo {author}
  {\bibfnamefont {K.}~\bibnamefont {Watanabe}}, \bibinfo {author}
  {\bibfnamefont {T.}~\bibnamefont {Taniguchi}}, \bibinfo {author}
  {\bibfnamefont {D.}~\bibnamefont {Shahar}}, \bibinfo {author} {\bibfnamefont
  {J.}~\bibnamefont {Hone}}, \bibinfo {author} {\bibfnamefont {C.~R.}\
  \bibnamefont {Dean}}, \bibinfo {author} {\bibfnamefont {A.~J.}\ \bibnamefont
  {Millis}}, \ and\ \bibinfo {author} {\bibfnamefont {A.~N.}\ \bibnamefont
  {Pasupathy}},\ }\href {\doibase 10.1038/s41567-019-0571-z} {\bibfield
  {journal} {\bibinfo  {journal} {Nature Physics}\ }\textbf {\bibinfo {volume}
  {15}},\ \bibinfo {pages} {947} (\bibinfo {year} {2019})}\BibitemShut
  {NoStop}%
\bibitem [{\citenamefont {Benyamini}\ \emph {et~al.}(2020)\citenamefont
  {Benyamini}, \citenamefont {Kennes}, \citenamefont {Telford}, \citenamefont
  {Watanabe}, \citenamefont {Taniguchi}, \citenamefont {Millis}, \citenamefont
  {Hone}, \citenamefont {Dean},\ and\ \citenamefont
  {Pasupathy}}]{Benyamini_2020}%
  \BibitemOpen
  \bibfield  {author} {\bibinfo {author} {\bibfnamefont {A.}~\bibnamefont
  {Benyamini}}, \bibinfo {author} {\bibfnamefont {D.~M.}\ \bibnamefont
  {Kennes}}, \bibinfo {author} {\bibfnamefont {E.~J.}\ \bibnamefont {Telford}},
  \bibinfo {author} {\bibfnamefont {K.}~\bibnamefont {Watanabe}}, \bibinfo
  {author} {\bibfnamefont {T.}~\bibnamefont {Taniguchi}}, \bibinfo {author}
  {\bibfnamefont {A.~J.}\ \bibnamefont {Millis}}, \bibinfo {author}
  {\bibfnamefont {J.}~\bibnamefont {Hone}}, \bibinfo {author} {\bibfnamefont
  {C.~R.}\ \bibnamefont {Dean}}, \ and\ \bibinfo {author} {\bibfnamefont
  {A.~N.}\ \bibnamefont {Pasupathy}},\ }\href {\doibase
  10.1021/acs.nanolett.0c04024} {\bibfield  {journal} {\bibinfo  {journal}
  {Nano Letters}\ } (\bibinfo {year} {2020}),\
  10.1021/acs.nanolett.0c04024}\BibitemShut {NoStop}%
\bibitem [{Note1()}]{Note1}%
  \BibitemOpen
  \bibinfo {note} {However, the Gauge function $\chi $ cannot be fixed uniquely
  only by this condition for the finite element method, i.e $\chi $ can only be
  determined up to a constant. Thus, we introduce another condition; $\DOTSI
  \intop \ilimits@ _S \Psi dS=0$, where $S$ is the entire sample domain.\cite
  {Qiang1994, fan2019}}\BibitemShut {NoStop}%
\bibitem [{\citenamefont {Carapella}\ \emph {et~al.}(2011)\citenamefont
  {Carapella}, \citenamefont {Sabatino},\ and\ \citenamefont
  {Costabile}}]{Carapella_2011}%
  \BibitemOpen
  \bibfield  {author} {\bibinfo {author} {\bibfnamefont {G.}~\bibnamefont
  {Carapella}}, \bibinfo {author} {\bibfnamefont {P.}~\bibnamefont {Sabatino}},
  \ and\ \bibinfo {author} {\bibfnamefont {G.}~\bibnamefont {Costabile}},\
  }\href {\doibase 10.1088/0953-8984/23/43/435701} {\bibfield  {journal}
  {\bibinfo  {journal} {Journal of Physics: Condensed Matter}\ }\textbf
  {\bibinfo {volume} {23}},\ \bibinfo {pages} {435701} (\bibinfo {year}
  {2011})}\BibitemShut {NoStop}%
\bibitem [{\citenamefont {Sabatino}\ \emph {et~al.}(2011)\citenamefont
  {Sabatino}, \citenamefont {Carapella},\ and\ \citenamefont
  {Costabile}}]{Sabatino_2011}%
  \BibitemOpen
  \bibfield  {author} {\bibinfo {author} {\bibfnamefont {P.}~\bibnamefont
  {Sabatino}}, \bibinfo {author} {\bibfnamefont {G.}~\bibnamefont {Carapella}},
  \ and\ \bibinfo {author} {\bibfnamefont {G.}~\bibnamefont {Costabile}},\
  }\href {\doibase 10.1088/0953-2048/24/12/125007} {\bibfield  {journal}
  {\bibinfo  {journal} {Superconductor Science and Technology}\ }\textbf
  {\bibinfo {volume} {24}},\ \bibinfo {pages} {125007} (\bibinfo {year}
  {2011})}\BibitemShut {NoStop}%
\bibitem [{Note2()}]{Note2}%
  \BibitemOpen
  \bibinfo {note} {When vortices move out of the finite sample, critical force
  which overcomes any barrier is needed. For a large enough slab, such an
  effect will be irrelevant since the collective force of the current over all
  vortices is much larger than any barrier.}\BibitemShut {Stop}%
\bibitem [{SM()}]{SM}%
  \BibitemOpen
  \href@noop {} {\ }\bibinfo {note} {See
  \href{https://doi.org/10.5281/zenodo.5802290}{Supplemental
  Material}}\BibitemShut {NoStop}%
\bibitem [{\citenamefont {Logg}\ \emph {et~al.}(2012)\citenamefont {Logg},
  \citenamefont {Mardal}, \citenamefont {Wells} \emph
  {et~al.}}]{LoggMardalEtAl2012a}%
  \BibitemOpen
  \bibfield  {author} {\bibinfo {author} {\bibfnamefont {A.}~\bibnamefont
  {Logg}}, \bibinfo {author} {\bibfnamefont {K.-A.}\ \bibnamefont {Mardal}},
  \bibinfo {author} {\bibfnamefont {G.~N.}\ \bibnamefont {Wells}},  \emph
  {et~al.},\ }\href {\doibase 10.1007/978-3-642-23099-8} {\emph {\bibinfo
  {title} {Automated Solution of Differential Equations by the Finite Element
  Method}}}\ (\bibinfo  {publisher} {Springer},\ \bibinfo {year}
  {2012})\BibitemShut {NoStop}%
\bibitem [{Note3()}]{Note3}%
  \BibitemOpen
  \bibinfo {note} {Specifically, both real and imaginary part of the complex
  order parameter $\Re [\Delta ]$ and $\Im [\Delta ]$ are set to be $\protect
  \sqrt {0.5}+\delta $, where $\delta $ is chosen randomly from a uniform
  distribution $\delta \in (-0.01, 0.01]$. However, since we concentrate on
  analyzing the steady state none of the general conclusions are affected by
  the initial conditions.}\BibitemShut {Stop}%
\bibitem [{\citenamefont {Machida}\ and\ \citenamefont
  {Kaburaki}(1993)}]{Machida1993}%
  \BibitemOpen
  \bibfield  {author} {\bibinfo {author} {\bibfnamefont {M.}~\bibnamefont
  {Machida}}\ and\ \bibinfo {author} {\bibfnamefont {H.}~\bibnamefont
  {Kaburaki}},\ }\href {\doibase 10.1103/PhysRevLett.71.3206} {\bibfield
  {journal} {\bibinfo  {journal} {Phys. Rev. Lett.}\ }\textbf {\bibinfo
  {volume} {71}},\ \bibinfo {pages} {3206} (\bibinfo {year}
  {1993})}\BibitemShut {NoStop}%
\bibitem [{\citenamefont {Embon}\ \emph
  {et~al.}(2017{\natexlab{b}})\citenamefont {Embon}, \citenamefont {Anahory},
  \citenamefont {Jeli{\'c}}, \citenamefont {Lachman}, \citenamefont
  {Myasoedov}, \citenamefont {Huber}, \citenamefont {Mikitik}, \citenamefont
  {Silhanek}, \citenamefont {Milo{\v{s}}evi{\'c}}, \citenamefont {Gurevich}
  \emph {et~al.}}]{embon2017imaging}%
  \BibitemOpen
  \bibfield  {author} {\bibinfo {author} {\bibfnamefont {L.}~\bibnamefont
  {Embon}}, \bibinfo {author} {\bibfnamefont {Y.}~\bibnamefont {Anahory}},
  \bibinfo {author} {\bibfnamefont {{\v{Z}}.~L.}\ \bibnamefont {Jeli{\'c}}},
  \bibinfo {author} {\bibfnamefont {E.~O.}\ \bibnamefont {Lachman}}, \bibinfo
  {author} {\bibfnamefont {Y.}~\bibnamefont {Myasoedov}}, \bibinfo {author}
  {\bibfnamefont {M.~E.}\ \bibnamefont {Huber}}, \bibinfo {author}
  {\bibfnamefont {G.~P.}\ \bibnamefont {Mikitik}}, \bibinfo {author}
  {\bibfnamefont {A.~V.}\ \bibnamefont {Silhanek}}, \bibinfo {author}
  {\bibfnamefont {M.~V.}\ \bibnamefont {Milo{\v{s}}evi{\'c}}}, \bibinfo
  {author} {\bibfnamefont {A.}~\bibnamefont {Gurevich}},  \emph {et~al.},\
  }\href@noop {} {\bibfield  {journal} {\bibinfo  {journal} {Nature
  communications}\ }\textbf {\bibinfo {volume} {8}},\ \bibinfo {pages} {1}
  (\bibinfo {year} {2017}{\natexlab{b}})}\BibitemShut {NoStop}%
\bibitem [{\citenamefont {Khusnatdinov}\ \emph {et~al.}(2000)\citenamefont
  {Khusnatdinov}, \citenamefont {Nagle},\ and\ \citenamefont
  {Nunes~Jr}}]{khusnatdinov2000ultrafast}%
  \BibitemOpen
  \bibfield  {author} {\bibinfo {author} {\bibfnamefont {N.}~\bibnamefont
  {Khusnatdinov}}, \bibinfo {author} {\bibfnamefont {T.}~\bibnamefont {Nagle}},
  \ and\ \bibinfo {author} {\bibfnamefont {G.}~\bibnamefont {Nunes~Jr}},\
  }\href@noop {} {\bibfield  {journal} {\bibinfo  {journal} {Applied Physics
  Letters}\ }\textbf {\bibinfo {volume} {77}},\ \bibinfo {pages} {4434}
  (\bibinfo {year} {2000})}\BibitemShut {NoStop}%
\bibitem [{wor()}]{work_progress}%
  \BibitemOpen
  \href@noop {} {\ }\bibinfo {note} {Takuya Okugawa, Avishai Benyamini, Andrew
  J. Millis, and Dante M. Kennes, work in progress.}\BibitemShut {Stop}%
\bibitem [{\citenamefont {Du}(1994)}]{Qiang1994}%
  \BibitemOpen
  \bibfield  {author} {\bibinfo {author} {\bibfnamefont {Q.}~\bibnamefont
  {Du}},\ }\href {\doibase 10.1080/00036819408840240} {\bibfield  {journal}
  {\bibinfo  {journal} {Applicable Analysis}\ }\textbf {\bibinfo {volume}
  {53}},\ \bibinfo {pages} {1} (\bibinfo {year} {1994})}\BibitemShut {NoStop}%
\bibitem [{\citenamefont {Fan}\ \emph {et~al.}(2019)\citenamefont {Fan},
  \citenamefont {Samet},\ and\ \citenamefont {Zhou}}]{fan2019}%
  \BibitemOpen
  \bibfield  {author} {\bibinfo {author} {\bibfnamefont {J.}~\bibnamefont
  {Fan}}, \bibinfo {author} {\bibfnamefont {B.}~\bibnamefont {Samet}}, \ and\
  \bibinfo {author} {\bibfnamefont {Y.}~\bibnamefont {Zhou}},\ }\href
  {https://projecteuclid.org:443/euclid.ojm/1554278424} {\bibfield  {journal}
  {\bibinfo  {journal} {Osaka J. Math.}\ }\textbf {\bibinfo {volume} {56}},\
  \bibinfo {pages} {269} (\bibinfo {year} {2019})}\BibitemShut {NoStop}%
\end{thebibliography}%

\end{document}